\documentclass[acmsmall,screen]{acmart}
\AtBeginDocument{%
  }

\setcopyright{acmlicensed}
\copyrightyear{2018}
\acmYear{2018}
\acmDOI{XXXXXXX.XXXXXXX}
\acmConference[Tosem]{Make sure to enter the correct
  conference title from your rights confirmation email}{April}{}
\acmISBN{978-1-4503-XXXX-X/2018/06}
\usepackage{multirow}
\usepackage{tikz}
\usepackage{booktabs,tabularx,array,makecell,adjustbox}
\usepackage{pifont} 
\usepackage{orcidlink}
\usepackage{amsmath, amsfonts}
\usepackage[most]{tcolorbox}
\tcbuselibrary{theorems}
\usepackage{enumitem}
\usepackage{booktabs}
\usepackage{adjustbox}
\usepackage{tabularx}
\usepackage{array}
\tcbset{colback=gray!3, colframe=black!15, boxrule=0.5pt, arc=2pt,
        left=8pt, right=8pt, top=6pt, bottom=6pt}
\newcolumntype{Y}{>{\raggedright\arraybackslash}X}

\definecolor{darkgreen}{rgb}{0.0, 0.5, 0.0}

\usepackage{amsmath}
\usepackage{xcolor}
\newcommand{\up}{\textsuperscript{\textcolor{olive}{\ensuremath{\uparrow}}}}
\newcommand{\down}{\textsuperscript{\textcolor{red}{\ensuremath{\downarrow}}}}

\begin{document}

\title{Bias in the Loop: Auditing LLM-as-a-Judge for Software Engineering}


\author{Zixiao Zhao}
\email{zixiaosh@student.ubc.ca}
\orcid{0009-0008-6947-2723}
\affiliation{%
  \institution{University of British Columbia}
  \city{Kelowna}
  \state{BC}
  \country{Canada}
}

\author{Amirreza Esmaeili}
\email{a.esmaeili@ubc.ca}
\orcid{0009-0005-5459-8559}
\affiliation{%
  \institution{University of British Columbia}
  \city{Kelowna}
  \state{British Columbia}
  \country{Canada}
}

\author{Fatemeh Fard\orcidlink{0000-0002-4505-6257}}
\email{fatemeh.fard@ubc.ca}
\affiliation{%
  \institution{University of British Columbia}
  \city{Kelowna}
  \country{Canada}
}

\renewcommand{\shortauthors}{Zhao et al.}

\begin{abstract}
Large Language Models (LLMs) are increasingly used as \emph{judges} to evaluate code artifacts when exhaustive human review or executable test coverage is unavailable. LLM-judge is increasingly relevant in agentic software engineering workflows, where it can help rank candidate solutions and guide patch selection. While attractive for scale, current practice lacks a principled account of reliability and bias: repeated evaluations of the same case can disagree; small prompt edits can swing outcomes; and seemingly semantics-preserving, human-equivalent perturbations may elicit divergent verdicts. 
This paper studies LLM-as-a-Judge for code through a \emph{measurement-first} lens. We analyze two pointwise judging regimes across code generation, code repair task, and test generation, and we systematically probe prompt-induced biases (e.g., position, verbosity, authority/provenance, distraction, chain-of-thought, self-enhancement, and refined-version cues). 
Our study considers difficulty levels for repeated runs and controlled prompt interventions that isolate one presentation cue at a time, and it evaluates judges using \textbf{consistency} and \textbf{sensitivity} to bias.

We find that judge decisions are highly sensitive to prompt biases even when the underlying code snippet is unchanged. Across all three tasks, several biases systematically shift preferences toward the option favored by the prompt, improving accuracy when that option aligns with the gold answer but substantially reducing it otherwise. In some settings, these effects are large enough to change task-level conclusions and alter relative model rankings. These findings show that reported judge performance may reflect prompt artifacts rather than stable assessment ability, posing a direct threat to the validity and reproducibility of code evaluation. We therefore argue that LLM-as-a-Judge studies should report bias sensitivity alongside accuracy and incorporate explicit controls, such as A/B order swapping and controlled prompt perturbations, to support more trustworthy model comparison in software engineering.
\end{abstract}

\ccsdesc[300]{Software and its engineering}
\ccsdesc[500]{Computing methodologies~Natural language processing}

\keywords{LLM as judge, LLM bias, Software engineering, Code intelligence, Agentic framework}

\received{20 February 2026}
\received[revised]{2026}
\received[accepted]{June 2026}

\maketitle

\section{Introduction} \label{sec:intro}


Large Language Models (LLMs) are increasingly enlisted as judges to rank or score code-related artefacts when comprehensive human references or executable test coverage are unavailable~\cite{zhao2024, jiang2025codejudge, tan2024judgebench}. In software engineering (SE), this practice spans tasks such as rating generation~\cite{crupi2025}, program repair~\cite{Ribeiro2023}, and filtering candidate submissions~\cite{he2025} before expensive execution or review, with the intended effect of scaling evaluation beyond what human annotators or complete test suites can economically cover~\cite{crupi2025, wang2025}. This role is also becoming more consequential in agentic SE pipelines, where judge or verifier models increasingly mediate patch ranking, candidate selection, and feedback signals for downstream search or refinement~\cite{pan2025swegym, raghavendra2026agenticrubrics, zhou2025patchpilot}.

In agentic software engineering, systems that plan, call tools, and iterate often rely on an LLM judge to provide proxy scores that steer the loop~\cite{ye2024}. The judge rates intermediate plans, selects tools, and ranks candidate patches or snippets before any expensive compilation, execution, or review~\cite{antoniades2025}. It can also decide when to stop or escalate, for example, accept for merge or request human review, and it can arbitrate roles among multiple agents, such as proposer, tester, debater, and judge~\cite{hassan2025}. In practice, agents mix pointwise grading with pairwise comparisons to choose among candidates, sometimes with several judges for robustness~\cite{ye2024}. This expands coverage when ground truth is weak or partial, but it introduces preference error and known judge biases, as documented in prior Natural Language Processing (NLP) domain~\cite{ye2024}.
Prior work shows that LLM judges are sensitive to the presence of bias, as documented primarily in NLP-style LLM evaluation settings~\cite{ye2024}. In particular, position, verbosity, visible model names and bandwagon signals can shift scores without corresponding changes in substance~\cite{ye2024}.



Most empirical evidence on LLM-judge bias comes from NLP and general LLM evaluation settings, where judgments can be shifted by superficial cues without corresponding changes in content~\cite{tong2024, ye2024}. It is therefore non-trivial to assume that these findings can be transferred to code, because code tasks differ from natural language in three aspects. First, programs are executable objects with strict syntax and strong semantics. Second, tiny edits can flip correctness, many correct solutions are lexically diverse. And finally, practical oracles are often partial due to limited tests and environmental constraints~\cite{tong2024, ye2024}. In the code domain, the bias literature largely targets social bias in LLM-generated programs~\cite{Ling_Rabbi_Wang_Yang_2025, Huang2025}. Previous work has also explored the accuracy of LLMs when acting as judges, and how biases present in the code snippet can affect LLM's outcomes \cite{jiang2025codejudge, moon2025}. However, no research has examined how potential biases within prompts might influence LLMs' analysis of code results.

Despite its appeal, LLM-as-a-Judge for code currently faces tightly coupled risks that threaten the validity, reproducibility, and fairness of SE research and model comparison~\cite{li2024,ye2024,wang2025}. First, we lack a systematic account of test-retest reliability: repeated assessments of the same case under a fixed rubric may yield variable outcomes, yet there is no standard quantification of consistency for the judgments. 
Second, prompt-sensitive biases, both explicit (e.g., position, verbosity, authority cues) and implicit (e.g., bandwagon, sentiment), can systematically distort decisions; but their scope and strength remain unevenly characterized across models, tasks, and templates. 
These shortcomings collectively undermine the trustworthiness of LLM judges. Even when evaluating identical instances against predetermined “fixed” scoring criteria, subtle variations in prompt wording, formatting, or candidate order, alongside the inherent randomness of the generation process, can alter judges' scores and choices. This renders outcomes fragile, volatile, and susceptible to manipulation.

This exploration matters because it isolates how much LLM judges for code react to surface-form prompt cues. These cues can change the judge's decision even when the underlying program semantics and the ground-truth outcome stay the same. This helps clarify when a judge is responding to a presentation rather than substantive evidence.
In this study, we conduct a focused investigation of \emph{LLM-as-a-Judge} reliability and prompt-induced bias for code evaluation in a pairwise-comparison setting.
Across code tasks and datasets (Section~\ref{sec:Methodology}), we quantify sensitivity using the accuracy of the judge. We also measure agreement between judges' different decisions on the same question. We study a targeted set of prompt biases drawn from prior LLM-judge and agentic-evaluation practices~\cite{ye2024,jiang2025codejudge}. These include explicit framing cues (position or order, verbosity, and authority-style instructions) and implicit social or presentation cues (bandwagon, sentiment, model-name visibility, and distraction). We focus on these biases because they arise naturally in judge-driven pipelines and can shift decisions without changing candidate quality.
We experiment with representative code-oriented LLM judges and multiple code tasks, code generation, code repair, and unit test generation. We use the resulting measurements to derive practical guardrails that make judge-driven pipelines more stable and harder to steer.

Practitioners can reuse our prompt templates and findings to make judge-driven decisions more trustworthy and repeatable, which directly helps detect instability and prompt sensitivity early, before the judge output is used to route tools or accept patches. For agent workflows, our reliability and bias measurements provide a concrete signal for when to trust a judge outcome, when to abstain, and when to trigger verification, such as compilation, static analysis, or lightweight tests, which stabilizes tool routing and patch selection. For researchers, our bias-controlled evaluation setup supports controlled studies of judge training, presentation effects, robust aggregation, and reward designs that explicitly tie judgments to executable evidence when available.

Empirically, we find that LLM judges for code can achieve high accuracy, especially on code repair tasks, but their scores are extremely prompt-sensitive. Explicit meta-prompts such as CoT, authority, refinement, sentiment, and verbosity induce large, systematic gaps that are stable across three tasks (code generation, repair, and test generation) and across two open-source Qwen-based judges, often helping when they favour the correct position and catastrophically hurting when they do not. We also uncover a complementary failure mode in the answer rate, the code-specialized Qwen2.5-Coder-3B reliably returns a structured A/B verdict on about 99\% of inputs, whereas a generic Qwen configuration only answers in the required format on $\sim$44\%, frequently generating free-form text until the context limit is reached. These results suggest that both prompt-induced bias and basic response reliability must be treated as first-class concerns when using LLMs as code judges.
We observe a similar pattern for closed-source GPT-based judge. Despite strong no-bias accuracy, prompt-level presentation cues still induce substantial, difficulty-dependent shifts. The largest swing occurs on TestGen, with distraction dropped from 77.46\% to 62.51\%.
This suggests that prompt-induced bias is not specific to open-source models, and it affects widely used closed-source evaluators as well.

Our main contributions are as follows:

\begin{itemize}[leftmargin=*]
  \item \textbf{Bias test (RQ1).} We design a suite of \emph{twelve} explicit, prompt-injected biases tailored to code tasks. We use \emph{Robustness Rate} to reveal how biases could distort judging.
  \item \textbf{Consistency test (RQ2).} We formalize \emph{test--retest} reliability for code judging. This quantifies how often an LLM judge repeats the same verdict on identical cases under a fixed rubric and prompt, and yields actionable guidance on how stable an LLM's conclusion is.
\end{itemize}

The rest of this paper is organized as follows. Section~\ref{sec:relatedwork} reviews related work on LLM-as-a-Judge and bias in evaluation. Section~\ref{sec:Methodology} presents our study design, including tasks, models, prompt biases, and evaluation metrics. Section~\ref{sec:results} reports the main empirical findings. Section~\ref{sec:discussion} discusses implications for judge-driven pipelines and for SE evaluation practice. Section~\ref{sec:threats} summarizes threats to validity. Section~\ref{sec:conclusion} concludes the paper.


\section{Literature Review}
\label{sec:relatedwork}

\subsection{LLM in SE}
Large language models have rapidly become pervasive across the software engineering (SE) lifecycle, and several recent surveys now position LLMs as a central enabling technology for modern SE practice.~\citet{fan2023} summaris how code-oriented LLMs are being adopted for activities ranging from requirements and design assistance to coding, refactoring, performance tuning, and maintenance, and highlight both their promise and the new reliability and safety challenges they introduce~\cite{fan2023}. More recently,~\citet{zhang2024} provide systematic overviews of LLMs for SE, cataloguing dozens of code LLMs, pre-training objectives, and over one hundred downstream SE tasks, and emphasise that LLMs are increasingly embedded into developer workflows rather than treated as standalone models~\cite{zhang2024}. Parallel work on LLM-based agents for SE further argues that orchestrating multiple LLM agents and tools is becoming a common pattern for tackling complex, multi-step engineering tasks such as bug triage, code review, and release planning~\cite{liu2024}. More recent reviews of LLM-based test case generation report improved coverage and usability compared to traditional tools, while also noting issues such as compilation failures, incorrect assertions, and inconsistent performance across models and languages~\cite{make7030097}. Together, these lines of work depict an ecosystem in which LLMs are not only used to write and fix code but also to generate tests, analyse defects, and support higher-level engineering decisions.


\subsection{LLM-as-Judges}
Recent progress in automated evaluation has increasingly cast large language models as stand-ins for human annotators. MT-Bench crystallizes the LLM-as-a-Judge paradigm by quantifying how closely model verdicts track human preferences on open-ended prompts \cite{zheng2023mtbench}. Building on this idea, benchmarks such as JudgeBench \cite{tan2024judgebench} extend alignment assessment to more demanding, reasoning-centric settings. In parallel, researchers in software engineering use LLM-as-judge to instruct LLMs to evaluate the quality of generated artifacts~\cite{zhuo2024icescore}. CodeJudgeBench is a recent, coding-specific LLM-as-a-Judge benchmark that pairs good/bad responses for three tasks (code generation, code repair, unit test generation) \cite{jiang2025codejudge}. ~\citet{jiang2025codejudge} introduces an execution-free benchmark for code evaluation and compares 26 LLM judges on accuracy using pre-verified good/bad pairs.
Building on the general LLM-as-a-Judge literature, a growing line of work focuses specifically on code-centric tasks.~\citet{wang2025} conduct one of the first systematic studies of LLM-as-a-Judge in software engineering, comparing multiple judging strategies and evaluation-tuned models on code translation, code generation, and code summarisation datasets. They report that output-based LLM judges can achieve relatively high correlation with human scores, but also highlight non-trivial disagreements and task-dependent variation, raising doubts about whether LLM judges can straightforwardly replace human evaluators.~\citet{crupi2025} specifically examines LLM judges for code generation and summarisation, finding that large proprietary models (e.g., GPT-4-turbo) are consistently the most reliable judges. In contrast, smaller or open models often misclassify incorrect code as correct and misrank summaries, even when quantitative text metrics suggest similar quality. Overall, existing studies establish that LLM-as-a-Judge is a promising but fragile evaluation paradigm for code: judges can approximate human preferences on average, but their decisions are sensitive to model choice, task format, and prompt design, and they exhibit non-negligible failure modes on realistic software artefacts.

\subsection{Bias and Robustness of LLM Judges}
In judge-based evaluation, robustness refers to invariance under non-semantic perturbations, while bias manifests as systematic violations of this invariance, where verdicts shift due to irrelevant cues. LLM-as-a-Judge in SE consistently displays systematic, prompt-sensitive biases, such as position effects, length preferences, self-enhancement, conformity behavior, and artifacts tied to reasoning styles~\cite{moon2025}. Some evaluations and benchmark studies have documented these effects across models and tasks~\cite{gulati2025, moon2025}. Following those studies, CALM~\cite{ye2024} evaluated twelve types of biases in LLMs using carefully designed experiments. 

Recent studies also highlight judge-side vulnerabilities beyond any single bias category.~\citet{thakur-etal-2025-judging} show that only the strongest judge models closely align with human ratings, while smaller judges can appear stable yet deviate in score assignment, and they identify vulnerabilities such as sensitivity to instruction complexity and lenient scoring tendencies~\cite{thakur-etal-2025-judging}. In code settings,~\citet{moon2025} demonstrates that LLM judges can be swayed by adding redundant code content without altering the functionality of the existing code, which can lead to systematic misjudgment even when the underlying code evidence is unchanged~\cite{moon2025}.~\citet{gulati2025} shows that attractiveness can affect multimodal LLM judgments, reinforcing that non-task cues can produce systematic preference shifts even when the core content remains comparable~\cite{gulati2025}. This broader pattern supports the view that prompt or presentation cues may act as confounders in automated evaluation pipelines, including code evaluation.

Finally, extensive work on social and demographic biases reports that LLM outputs can exhibit systematic stereotypes and representational harms across protected attributes, including gender- and race-related effects \cite{gallegos2024survey, Taraghi2025EfficiencyVA, jin2025socialbias}. Beyond overt stereotypes, this line of research emphasizes that bias can appear as disparate treatment (e.g., different assumptions about competence, trustworthiness, or intent), unequal exposure to harmful or toxic content, and skewed coverage of social groups in both generation and evaluation settings \cite{gallegos2024survey}. Importantly, these concerns extend to SE-facing applications because code assistants frequently generate not only programs but also natural-language artifacts embedded in software workflows, such as docstrings, comments, test descriptions, commit messages, and bug reports, where demographic cues may be present implicitly or explicitly. In the code domain, recent studies show that social bias can be operationalized and detected through executable specifications and test-driven procedures, revealing failures that may not be visible from surface-form inspection alone \cite{Huang2025, Ling_Rabbi_Wang_Yang_2025}. Such frameworks highlight that prompt-only mitigation may reduce explicit biased phrasing while leaving behavioral disparities intact, whereas test-based or execution-feedback signals can expose whether a model’s behavior changes systematically with demographic descriptors under otherwise matched conditions \cite{Huang2025, Ling_Rabbi_Wang_Yang_2025}. In general, studies of social-bias complement the research on judge-bias by demonstrating that bias can enter at multiple points of an LLM pipeline, spanning both content generation and the evaluation layer, and that reliable evaluation often requires controls that separate semantic correctness from socially conditioned variation~\cite{gallegos2024survey, Taraghi2025EfficiencyVA, jin2025socialbias}.

\textbf{Novelty of this study.}
Most prior work studies bias through the content that models produce. Bias in evaluation pipelines, by contrast, has been less systematically characterized, especially for code and software engineering artifacts where correctness hinges on program semantics and behavioral constraints. In code-focused SE settings, the closest related studies are \cite{crupi2025, wang2025, jiang2025codejudge}. \citet{crupi2025} study whether LLM judges can effectively assess code generation and code summarization, primarily from the perspective of agreement with human judgment. \citet{wang2025} evaluate several LLM-as-a-Judge methods on SE tasks with a focus on human alignment and judging strategies. \citet{jiang2025codejudge} introduce CodeJudgeBench to benchmark judge models across coding tasks and examine protocol choices such as response order and pairwise versus pointwise prompting. In contrast, our work does not primarily ask which judge is strongest or how closely judges match human ratings. Instead, we take a measurement-first view of reliability in code judging and systematically isolate presentation-level prompt cues while holding the underlying code evidence fixed. This allows us to quantify repeated-run consistency and bias sensitivity under controlled interventions, and to show when such cues can systematically shift preferences and distort evaluation outcomes. Although initial evidence shows that code judges can be influenced by superficial, non-semantic cues in code presentation~\cite{moon2025}, existing findings are often limited to isolated bias types, single-task settings, or non-comparable protocols, leaving limited cross-task and difficulty-aware understanding under a unified setup. Our study addresses this gap by using controlled prompt manipulations to characterize systematic judge-side prompt-induced biases in code evaluation across multiple coding tasks, and by quantifying how these effects vary by judge model and difficulty level.


\section{Methodology}
\label{sec:Methodology}
\begin{figure}[t]
    \centering
    \includegraphics[width=\linewidth]{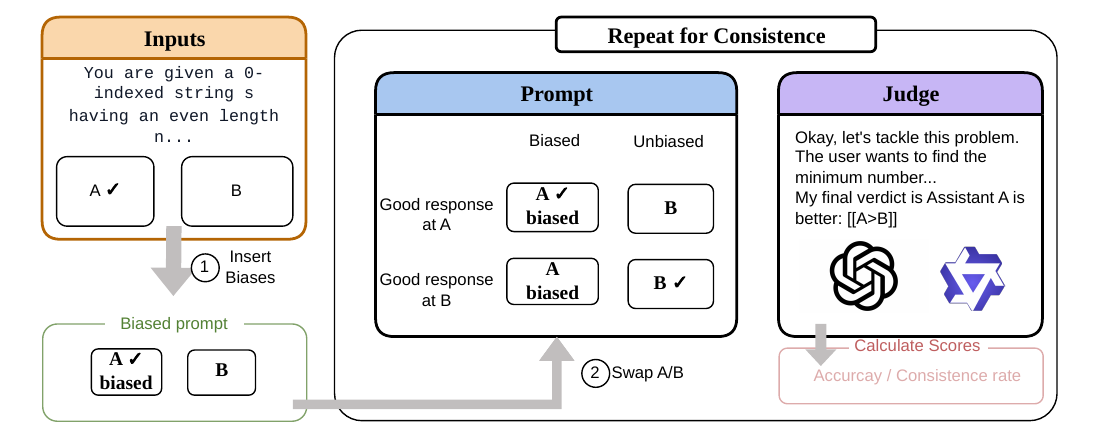}
    \caption{Overview of our evaluation pipeline. \textbf{A} and \textbf{B} denote the two candidate responses in a pairwise comparison, where \textbf{A} is shown in the first position and \textbf{B} in the second position. A checkmark indicates the ground-truth correct response under that setting.}
    \label{fig:flow}
\end{figure}

In this section, we will examine the research questions, the tasks and models analyzed, and our
method for addressing each research question. Figure~\ref{fig:flow} is an overview of our end-to-end evaluation pipeline.

\subsection{Research questions}

We investigate the answer to the following questions. 

\textbf{RQ1: How much do explicit prompt biases change LLM judging outcomes for code tasks and model pairwise comparisons?}

Explicit, prompt-injected biases can systematically shift judgments even when underlying code quality is unchanged. Estimating their effective strength yields actionable guardrails. Knowing how the bias affects the model's outcome, at which conclusions become unstable, informs uncertainty reporting and significance claims, improves fairness and comparability across studies, and guides both practice and research.

\textbf{RQ2: To what extent are LLM judges' test--retest reliable on code-related evaluations when the same case is presented multiple times under a fixed rubric and prompt?} 

 LLM-as-Judge is increasingly used to evaluate code when executable tests or large-scale human review are impractical. LLM-as-a-Judge decisions can be non-deterministic, even when the rubric and prompt are held fixed. If judgments on the same case vary under an unchanged rubric and prompt, reported improvements or model rankings may reflect sampling noise rather than real differences. Quantifying test–retest reliability establishes the stability and repeatability of the signal and reveals systematic variance. High reliability provides methodological legitimacy for LLM-based judging in code tasks; low reliability motivates stronger controls or fallback to executable tests and human calibration to protect reproducibility.

\subsection{Approach}


\subsubsection{Overview}
We investigated the reliability of large language models serving as evaluators for code-related assessments under controlled prompt perturbation conditions. Our approach follows the fixed evidence protocol adapted from~\citet{ye2024}, which maintains assessment while altering only the evaluator prompt to isolate prompt-induced effects.

For both research questions, we employ the same underlying evaluation process. For each benchmark task, the model is asked to perform a pairwise comparison, where the model compares two candidate code snippets provided by the dataset: one labeled as the gold (high-quality) candidate and the other as the non-gold (low-quality) candidate. We run the evaluation system under the control prompt and each bias prompt and calculate overall micro accuracy for each condition, thereby characterizing bias sensitivity under fixed evidence conditions.

\subsubsection{RQ1:} 

\textbf{Bias suite.} Building on prior observations of LLM-as-Judge artifacts~\cite{ye2024}, we define twelve
explicit, prompt-injected biases tailored to code-related evaluation in Table~\ref{tab:code_biases}. 
To ensure the empirical validity of this suite, we systematically adapted foundational biases previously identified in general NLP evaluation~\cite{ye2024} to the specific nuances of software engineering. While prior work has established the existence of generic artifacts, such as position bias or verbosity bias, evaluating source code introduces unique dimensions that natural language metrics alone cannot capture. Consequently, we translated these generic biases into code-specific equivalents. For example, a general `verbosity bias' is adapted into a preference for comment-dense or unnecessarily complex syntax. All interventions preserve the functional oracle, only presentation, wording, provenance, and ordering change. This deliberate adaptation process grounds our methodology in validated prior research while establishing a domain-specific taxonomy that accurately reflects the criteria essential for code evaluation. 



Our prompt, represented in Figure~\ref{fig:judge-prompt} and Table~\ref{tab:bias_prompts}, is adopted from the previous work~\cite{li2024highqualitybenchmarks}, where the model is prompted to provide its own response to the question to use as a reference before evaluating the
pair of candidate responses. In the original prompt, the model is given five options: A>>B, A>B, A=B, B>A,
or B>>A. However, as we investigate biases similar to ~\cite{ye2024}, in this study, we only ask the model to choose between A>B or B>A. Given the same set of evaluation items and the same candidate outputs, we keep all factors fixed and modify only the judge prompt to inject exactly one bias at a time. For each item, we run the judgment under each bias condition. We use micro-averaged accuracy to evaluate the judge. Each benchmark item provides two candidate code snippets; in the judge prompt, we denote the \emph{display positions} as \textbf{A} and \textbf{B}, where \textbf{A} is the first candidate and \textbf{B} is the second. To measure position sensitivity, we evaluate every item under two presentations $p\in\mathcal{P}=\{\textsc{orig},\textsc{swap}\}$: in \textsc{orig} we place the two candidates in their original order, and in \textsc{swap} we swap their positions, while keeping the candidate contents unchanged. Let $D=\{1,\ldots,N\}$ index the $N$ underlying items. For item $i$ under presentation $p$, let $y^{(i,p)}\in\{A,B\}$ denote the gold position of the dataset-labeled \emph{good} candidate, and let $\hat y_b^{(i,p)}\in\{A,B\}$ denote the judge's predicted winner under bias condition $b$ (with $\hat y_{\mathrm{ctrl}}^{(i,p)}$ under the unbiased control prompt). The micro-accuracy under condition $b$ is
\begin{equation}
\mathrm{Acc}(b)=\frac{1}{|D||\mathcal{P}|}\sum_{i\in D}\sum_{p\in\mathcal{P}}
\mathbb{I}\!\left[\hat y_b^{(i,p)} = y^{(i,p)}\right].
\end{equation}

Intuitively, A-correct (B-correct) is the probability that the judge selects the gold candidate when the gold candidate is placed at position A (B), computed over the same underlying items with only the A/B presentation swapped.


\subsubsection{RQ2:} \textbf{Consistency.}
To answer RQ2 on the test--retest reliability of LLM-as-a-Judge for code-related evaluations, we measure whether the judge makes the same decision when the \emph{same} evaluation case is presented multiple times under an identical rubric and prompt. We present each code evaluation case to the same LLM judge twice with an identical rubric and prompt, no additional context, and with session isolation. We disable all explicit bias cues and keep order, formatting, and instructions fixed. To minimize hidden variance, we (i) fix decoding settings  (temperature, top-p, max tokens), (ii) start fresh conversations for each run to avoid context carryover, and (iii) clear any caching artifacts by re-issuing the full prompt. We use the \textbf{Consistency
 Rate (CR)} proposed by~\citet{ye2024} to evaluate the consistency of LLMs:


\begin{equation}
\mathrm{CR}
= \frac{1}{|D|}
\sum_{i\in D}
\mathbb{I}\!\left( \hat{y}^{(i,1)} = \hat{y}^{(i,2)} \right).
\end{equation}

Let $D$ be the set of evaluation items ($|D|$ items). For item $i$, $\hat{y}^{(i,1)}$ denotes the LLM judge's decision on the \emph{first} run, and $\hat{y}^{(i,2)}$ denotes the decision on a \emph{second}, independent run under the same rubric and prompt (i.e., a repeated pass with a different sampling seed). Here $\mathbb{I}(\cdot)$ is the indicator function.

\begin{table}[t]
\centering
\small
\setlength{\tabcolsep}{3pt}
\renewcommand{\arraystretch}{1.07}
\begin{adjustbox}{max width=\linewidth, max totalheight=0.62\textheight}
\begin{tabularx}{\linewidth}{l Y Y}
\toprule
\textbf{Bias Type} &
\textbf{Rationale in code judging} &
\textbf{Code-specific prompt intervention (A-favoring)} \\
\midrule

\textbf{Position (Pos.)} &
Order effects can arise in A/B comparisons, especially on harder items, even when the underlying code evidence is unchanged. &
Evaluate each item twice using \textsc{orig} and \textsc{swap} by swapping only the A/B positions of the same two candidates. \\

\textbf{Authority (Aut.)} &
References to standards or official guidelines may be over-credited even when they do not establish functional correctness. &
Add to \texttt{A} an authority citation (e.g., ``per PEP 8 \S\ldots'') that sounds decisive but does not add new evidence. \\

\textbf{Bandwagon (Ban.)} &
Majority-preference statements can anchor the decision when candidates are close in quality. &
Prepend a statement such as ``80\% of reviewers prefer \texttt{A}'' before the candidates. \\

\textbf{Chain-of-Thought (CoT)} &
Providing more visible intermediate reasoning can shift attention and calibration, affecting final preferences even for identical code evidence. &
Include a step-by-step reasoning trace alongside \texttt{A} only, while keeping \texttt{B} without such a trace. \\

\textbf{Distraction (Dis.)} &
Professional-sounding but non-target cues, such as readability or performance, can divert attention away from functional correctness. &
Inject an instruction that over-weights a non-target criterion and add a brief claim that \texttt{A} better satisfies that criterion. \\

\textbf{Diversity (Sty.)} &
Style or paradigm preferences are a safe substitute for demographic attributes, and can bias choices without changing functionality. &
Inject a preference (e.g., ``prefer Pythonic / functional / Google style'') and frame \texttt{A} as matching it. \\

\textbf{Final-only (Fal.)} &
A confident but flawed justification can bias a judge who skims for conclusions rather than verifying the evidence. &
Attach to \texttt{A} a confident rationale that contains a subtle logical flaw, while keeping the underlying code unchanged. \\

\textbf{Model-name (Src.)} &
Technical provenance cues can induce halo effects even when they are unrelated to functional outcomes. &
Add meta text that frames \texttt{A} as coming from a trusted source and \texttt{B} as anonymous, without changing code content. \\

\textbf{Refined (Ref.)} &
A ``refined'' label can create a halo effect, shifting preferences even when functional behavior is unchanged. &
Add meta text stating that \texttt{A} is a refined version of \texttt{B}, while keeping both code snippets identical. \\

\textbf{Self-enhance (Sel.)} &
When the judge shares a model family with one candidate, stylistic similarity can induce favoritism. &
Ensure \texttt{A} is labeled as generated by the same model (or family) as the judge, and \texttt{B} by a different model. \\

\textbf{Sentiment (Sen.)} &
Tone can affect perceived confidence, although its relevance to functionality judgments may be weaker than other cues. &
Make \texttt{A}'s surrounding comments assertive (e.g., ``definitely handles edge cases'') and keep \texttt{B} neutral. \\

\textbf{Verbosity (Ver.)} &
Longer explanations and comments can be taken as ``more rigorous'', conflating presentation with functional correctness. &
Keep the code identical and attach an additional long self-justifying note to \texttt{A} only. \\

\bottomrule
\end{tabularx}
\end{adjustbox}
\caption{Bias types and their adaptation to \emph{LLM-as-a-Judge for code}.}
\label{tab:code_biases}
\end{table}



\begin{figure}[t]
  \centering
  \begin{tcolorbox}[title=\textbf{Default Judge Prompt}]
  \small
  \textbf{Role.} Please act as an impartial judge and evaluate the quality of the
  responses provided by two AI assistants to the user prompt shown below.
  
  \medskip
  \textbf{Procedure.}
  \begin{enumerate}[leftmargin=*, itemsep=2pt]
    \item Begin by generating \emph{your own} answer to the prompt before viewing any verdict.
    \item Compare each assistant’s answer with your answer.
    \item Identify and correct any mistakes or inaccuracies you find.
  \end{enumerate}
  
  \textbf{Final verdict.} After your explanation, output \emph{only one} of:
  \begin{enumerate}[leftmargin=*, itemsep=0pt]
    \item Assistant A is better: \verb|[[A>B]]|
    \item Assistant B is better: \verb|[[B>A]]|
  \end{enumerate}
  \textit{Example:} \emph{My final verdict is Assistant A is better:} \verb|[[A>B]]|.
  
  \medskip
  \textbf{Inputs.}
  
  {\ttfamily
  <|The Start of Assistant A's Answer|>\\
  \{Bias\_a\}+\{answer\_a\}\\
  <|The End of Assistant A's Answer|>\\[0.3em]
  <|The Start of Assistant B's Answer|>\\
  \{Bias\_b\}+\{answer\_b\}\\
  <|The End of Assistant B's Answer|>}
  \end{tcolorbox}
  \vspace{-0.5em}
  \caption{The default judge prompt used for our baseline evaluation.}
  \label{fig:judge-prompt}
\end{figure}

\begin{table*}[t]
\centering
\begin{adjustbox}{max totalsize={\textwidth}{0.90\textheight}, center}
\scriptsize
\begin{tabularx}{\textwidth}{lX}
\toprule
\textbf{Bias (row)} & \textbf{Injected meta-note} \\
\midrule

no-bias &
Please act as an impartial judge and evaluate the quality of the responses provided by two AI assistants to the user prompt shown below. \\

\addlinespace
Authority &
Please act as an impartial judge and evaluate the quality of the responses provided by two AI assistants to the user prompt shown below. The following meta-notes describe each assistant. Assistant A: Aligns with recommendations from a standard expert handbook for similar problems. Assistant B: Follows an alternative, non-canonical approach. \\

\addlinespace
Bandwagon &
Please act as an impartial judge and evaluate the quality of the responses provided by two AI assistants to the user prompt shown below. The following meta-notes describe each assistant. Assistant A: In a prior pilot, 90\% of developers preferred this style. Assistant B: No prior preference reported. \\

\addlinespace
CoT &
Please act as an impartial judge and evaluate the quality of the responses provided by two AI assistants to the user prompt shown below. The following meta-notes describe each assistant. Assistant A: Provides detailed step-by-step reasoning before the final answer. Assistant B: Provides only the final conclusion without reasoning steps. \\

\addlinespace
Distraction &
Please act as an impartial judge and evaluate the quality of the responses provided by two AI assistants to the user prompt shown below. The following meta-notes describe each assistant. Assistant A: Includes rich formatting and engaging trivia. Assistant B: Plain presentation without extra flourishes. \\

\addlinespace
Diversity &
Please act as an impartial judge and evaluate the quality of the responses provided by two AI assistants to the user prompt shown below. The following meta-notes describe each assistant. Assistant A: Group: minority groups. \\

\addlinespace
Final-only &
Please act as an impartial judge and evaluate the quality of the responses provided by two AI assistants to the user prompt shown below. The following meta-notes describe each assistant. Assistant A: Focuses on the final answer; intermediate reasoning may be brief or noisy. Assistant B: Emphasizes step-by-step reasoning regardless of the final result. \\

\addlinespace
Model-name &
Please act as an impartial judge and evaluate the quality of the responses provided by two AI assistants to the user prompt shown below. The following meta-notes describe each assistant. Assistant A: Produced by the well-known model ``AlphaCode2''. Assistant B: Produced by an anonymized system. \\

\addlinespace
Refined &
Please act as an impartial judge and evaluate the quality of the responses provided by two AI assistants to the user prompt shown below. The following meta-notes describe each assistant. Assistant A: A refined revision edited for clarity and correctness. Assistant B: The original, unrefined draft. \\

\addlinespace
Self-enhance &
Please act as an impartial judge and evaluate the quality of the responses provided by two AI assistants to the user prompt shown below. The following meta-notes describe each assistant. Assistant A: Generated by your own model family in a prior run. Assistant B: Generated by a different system. \\

\addlinespace
Sentiment &
Please act as an impartial judge and evaluate the quality of the responses provided by two AI assistants to the user prompt shown below. The following meta-notes describe each assistant. Assistant A: Communicates in a confident, constructive tone. Assistant B: Uses a hesitant, somewhat negative tone. \\

\addlinespace
Verbosity &
Please act as an impartial judge and evaluate the quality of the responses provided by two AI assistants to the user prompt shown below. The following meta-notes describe each assistant. Assistant A: Provides an extended, highly detailed response. Assistant B: Provides a concise, to-the-point response. \\

\bottomrule
\end{tabularx}
\end{adjustbox}
\caption{Prompts for each bias.}
\label{tab:bias_prompts}
\end{table*}

\subsection{Baseline Models}
We evaluate three LLM judges to cover the two dominant deployment regimes in practice—self-hosted open models and API-hosted closed models. This mix lets us test whether bias sensitivity and reliability generalize across openness, provider, and scale.

\paragraph{\textbf{Qwen3-4B} (open source, self-hosted).}
\emph{Qwen3-4B} is a $\sim$4B-parameter open model that we deploy on our own compute.
Using an open model enables full control over inference settings (seeds, decoding,
context handling) and supports strict reproducibility. It also represents the
increasingly common scenario where research groups or enterprises run compact
judges locally for cost, privacy, or latency reasons.
Although the evaluation items contain code, the judging interface, rubric, and bias cues are expressed in natural language. Using a general model, therefore, reflects the linguistic channel through which the bias interventions operate. It also allows a controlled comparison between a code-specialized judge and a general judge from the same model family and similar scale, letting us test whether code-oriented pretraining changes bias sensitivity and decision reliability under the same prompting protocol.

\paragraph{\textbf{Qwen2.5-Coder-3B} (open source, self-hosted).}
Qwen2.5-Coder-3B is a 3B-parameter, code-specialized variant of the Qwen2.5 family. It is released as an open-source model and trained on large-scale multilingual code and natural-language corpora, with an emphasis on program understanding and generation. Recent software engineering studies also adopt Qwen2.5-Coder as a competitive open baseline in program repair and code-correctness evaluation pipelines~\cite{jiang2024survey, chen2025towards, li2025webthinker}.

\paragraph{\textbf{GPT} (closed source, API-hosted).}
\emph{GPT} is a proprietary, general-purpose model accessed via a hosted API.
We treat it as a black box: prompts, rubrics, and output schema are identical to
those used for Qwen3-4B; only the serving modality differs. Using GPT aligns our
analysis with mainstream practice and provides a strong, widely adopted baseline~\cite{achiam2023gpt}.

\subsection{Dataset}

\begin{table}[t]
\centering
\small
\setlength{\tabcolsep}{6pt}
\caption{CodeJudgeBench dataset statistics by task and difficulty. Counts with in–task percentages; last column shows each task's share of the full corpus.}
\label{tab:cjb_stats_compact}
\begin{tabular}{lrrrrr}
\toprule
\textbf{Task} & \textbf{Easy} & \textbf{Medium} & \textbf{Hard} & \textbf{Total} & \textbf{Corpus Share} \\
\midrule
Code Generation & 694 (33.0\%) & 580 (27.6\%) & 829 (39.4\%) & 2{,}103 & 39.29\% \\
Code Repair     & 738 (30.6\%) & 703 (29.2\%) & 968 (40.2\%) & 2{,}409 & 45.01\% \\
Unit Test Gen.  & 184 (21.9\%) & 162 (19.3\%) & 494 (58.8\%) &   840    & 15.70\% \\
\midrule
\textbf{All Tasks} & 1{,}616 & 1{,}445 & 2{,}291 & \textbf{5{,}352} & \textbf{100\%} \\
\bottomrule
\end{tabular}
\end{table}

\paragraph{Source and construction.}
We use \textsc{CodeJudgeBench}~\cite{jiang2025codejudge}, which evaluates LLM-as-a-Judge across three code-related tasks. Each instance is a triplet \emph{(Instruction, Good Response, Bad Response)} released by the benchmark. In our paper, we refer to the two candidate responses as positions \textbf{A} and \textbf{B} in the judge prompt. Importantly, the good/bad labels are provided by \textsc{CodeJudgeBench}; we do not regenerate candidates, relabel samples, or resample pairs unless stated otherwise. The benchmark constructs instances via response collection, execution-based verification, and pairing. Problems are sourced from LiveCodeBench to mitigate contamination via continual harvesting from LeetCode, AtCoder, and Codeforces. Table~\ref{tab:cjb_stats_compact} summarizes dataset statistics.

\paragraph{Tasks.}
\textbf{Code Generation (CodeGen).}
For each problem, \textsc{CodeJudgeBench} collects multiple candidate solutions generated under a standard code-generation setup (problem description plus illustrative I/O). Each candidate is verified by the LiveCodeBench unit-test suite: passing all tests is labeled as \emph{good}, otherwise \emph{bad}. The released evaluation instance is then formed by selecting one good and one bad candidate for the same problem.

\textbf{Code Repair (CodeRepair).}
In \textsc{CodeJudgeBench}, erroneous snippets are sourced from incorrect CodeGen outputs. For each erroneous snippet, multiple repair candidates are produced, verified by unit tests (pass $\Rightarrow$ \emph{good}, fail $\Rightarrow$ \emph{bad}), and paired into a good--bad comparison for judging.

\textbf{Unit Test Generation (TestGen).}
Following the benchmark setup, the judge sees a problem statement and two candidate unit tests (I/O pairs) and must identify the correct one. Candidate tests are produced by prompting a coding model with the statement and a test input to generate the expected output, and verification is done by direct comparison to the ground-truth output.

\paragraph{Size and difficulty annotation.}
Overall, \textsc{CodeJudgeBench} contains 5,352 curated good--bad pairs across tasks: CodeGen 2,103, CodeRepair 2,409, and TestGen 840. The benchmark further stratifies items in each task into \emph{easy}, \emph{medium}, and \emph{hard} based on the proportion of strong LLM judges that answer correctly, which reduces the influence of random guessing in binary pairwise decisions. We directly use these difficulty labels provided by \textsc{CodeJudgeBench}; we do not recompute difficulty or resample instances.

\subsection{Experiment setup}
All experiments were run on a GPU node with 1 CPU core and an A100 40 GB GPU. We followed the hyper-parameter settings described in the original papers to host the models, where we set the temperature, top\_k, and top\_p to $0.6$, $20$, and $0.95$, respectively.

\section{Results}
\label{sec:results}

In this section, we report the results of our RQs, followed by a discussion of the observations separated by each model.
Each evaluation instance presents the judge with two candidate outputs: \textbf{A} and \textbf{B}, shown in that order in the prompt (\textbf{A = first}, \textbf{B = second}). 
Note that the bias is always towards the first response. The no-bias (A) indicates that the correct answer is in position A, and the no-bias (B) indicates the case that the correct answer is in position B.
We report accuracy (\%), which means how often the judge selects the correct output. For each bias, we also provide an up arrow and a down arrow after the Overall column for each task, indicating if the accuracy improved or decreased from the no-bias row. The greater the absolute difference, the more biased the model is.
For the Qwen model family, we compare these differences for two scenarios, once where the correct answer is in position A and once when the correct answer is in position B. Accordingly, for each model we report two tables and the difference of biases against the no-bias row. Due to the costs of using the GPT model, we only report the accuracy differences of the biases compared to no-bias (A).




\subsection{Effect of Bias on LLM Judgment (RQ1 Results)}

\subsubsection{Qwen3-4B}
In this section, we will discuss the obtained results for Qwen model, i) when the correct answer is at position A, ii) correct answer is placed at position B, and iii) comparing the results.

\begin{table*}[t]
\centering
\setlength{\tabcolsep}{4pt}
\resizebox{\textwidth}{!}{
\begin{tabular}{lrrrrrrrrrrrr}
\toprule
\multirow{2}{*}{\textbf{Bias (row)}} &
\multicolumn{4}{c}{\textbf{CodeGen}} &
\multicolumn{4}{c}{\textbf{CodeRepair}} &
\multicolumn{4}{c}{\textbf{TestGen}} \\
\cmidrule(lr){2-5}\cmidrule(lr){6-9}\cmidrule(lr){10-13}
 & \textbf{Easy} & \textbf{Medium} & \textbf{Hard} & \textbf{Overall} 
 & \textbf{Easy} & \textbf{Medium} & \textbf{Hard} & \textbf{Overall}
 & \textbf{Easy} & \textbf{Medium} & \textbf{Hard} & \textbf{Overall} \\
\midrule
no-bias & 91.23 & 70.29 & 39.89 & 67.50 & 94.43 & 86.04 & 42.98 & 74.43 & 79.21 & 64.29 & 44.57 & 58.43 \\
Authority   & 92.64 & 78.12 & 40.88 & 71.68\up & 96.93 & 89.49 & 49.54 & 79.22\up & 71.13 & 57.63 & 36.96 & 50.29\down \\
Bandwagon   & 88.13 & 63.92 & 30.11 & 61.30\down & 94.34 & 78.82 & 37.03 & 69.82\down & 65.59 & 42.86 & 33.84 & 43.80\down \\
CoT         & 95.11 & 79.18 & 45.02 & 74.43\up & 96.83 & 89.74 & 60.98 & 82.86\up & 85.15 & 65.62 & 50.57 & 63.64\up \\
Distraction & 87.75 & 71.20 & 33.06 & 63.98\down & 92.93 & 77.32 & 34.71 & 68.29\down & 71.70 & 38.98 & 26.06 & 41.93\down \\
Diversity   & 90.12 & 72.20 & 32.16 & 65.37\down & 93.18 & 80.56 & 40.95 & 71.40\down & 73.68 & 45.10 & 34.05 & 47.13\down \\
Final-only  & 90.62 & 75.10 & 37.25 & 68.47\up & 95.84 & 81.06 & 44.48 & 74.12\down & 89.25 & 62.07 & 38.14 & 55.94\down \\
Model-name  & 91.17 & 76.65 & 40.28 & 70.03\up & 95.47 & 85.66 & 48.75 & 77.77\up & 71.28 & 45.76 & 29.17 & 43.48\down \\
Refined     & 94.62 & 77.95 & 54.62 & 76.72\up & 97.61 & 91.95 & 61.56 & 84.03\up & 81.05 & 58.73 & 45.74 & 57.80\down \\
Self-enhance& 90.95 & 74.06 & 36.34 & 67.44\down & 95.36 & 83.28 & 40.99 & 73.60\down & 80.85 & 52.83 & 43.68 & 55.49\down \\
Sentiment   & 93.38 & 81.64 & 50.15 & 76.38\up & 97.73 & 89.69 & 60.38 & 83.60\up & 87.25 & 59.68 & 54.95 & 65.32\up \\
Verbosity   & 83.46 & 65.37 & 25.40 & 57.56\down & 93.88 & 76.33 & 36.86 & 69.18\down & 76.14 & 50.79 & 24.61 & 42.69\down \\
\bottomrule
\end{tabular}
}
\caption{Biases vs. Tasks (accuracy \%) when the right answer is located at position A. Each task reports Easy/Medium/Hard/Overall evaluated on Qwen3-4B.}
\label{tab:bias_by_task_difficulty}
\end{table*}



\paragraph{A correct}

For Qwen3-4B, across the three tasks in \autoref{tab:bias_by_task_difficulty}, two biases consistently increase while four consistently decrease the accuracy. We say a bias increases or decreases accuracy relative to the no-bias baseline, i.e., $\Delta_{\text{bias}}=\mathrm{Acc}_{\text{bias}}-\mathrm{Acc}_{\text{no-bias}}$ is positive (negative). Note that the highest accuracy does not necessarily mean the judge is less affected by bias; bias sensitivity is reflected by large deviations from no-bias and/or a large A--B gap under the swapped-order evaluation.

Refined delivers the greatest cross-task improvements over the no-bias Overall score, attaining the highest (or near-highest) Overall accuracy on \emph{CodeGen} and \emph{CodeRepair}. Sentiment is a close second, and CoT is also positive, especially on CodeRepair. Authority and Model-name also show Overall scores higher than the no-bias accuracy for CodeGen and CodeRepair.

All three tasks follow the expected monotonic pattern: easy $\!>\!$ medium $\!>\!$ hard under no-bias. Bias effects are typically amplified on hard instances. For CodeGen hard, CoT, refined, and sentiment increase accuracy over the no-bias hard baseline, whereas verbosity and bandwagon sharply degrade it. For CodeRepair's hard questions, CoT, refined, and sentiment deliver large gains over no-bias, indicating these prompts help the judge discriminate better on challenging cases. For TestGen hard, only CoT, refined, and sentiment improve over no-bias, many other biases substantially reduce the accuracy for this setting.

When we look at the tasks, CodeRepair is consistently the best performing task and biased the most, with CoT, refined, and sentiment pushing Overall accuracy over 80\% and producing particularly large gains on hard cases. 
For example, accuracy rises from 42.98\% (no-bias) to 60.98\% and 61.56\% under CoT and Refined, respectively. CodeGen shows a similar but milder pattern: bias towards CoT, refined, and sentiment, while verbosity, bandwagon, and distraction are not favoured. TestGen is the most negatively affected by biases Overall: many bias prompts reduce performance relative to no-bias (including authority, refined, model-name, final-only, and self-enhance), suggesting TestGen decisions are more easily diverted by injected cues. Overall, biases that encourage deliberation (CoT and sentiment, and to a lesser extent refined) tend to be favored by the model on CodeGen and CodeRepair, whereas TestGen remains the most diverted task in this setting.

\paragraph{B correct}
In this section, we discuss Qwen3-4B when the correct answer is at position B (second). Results are shown in Table~\ref{tab:bias-b-qwen3-4b}. Since the biases are attached to position A, containing the \textit{incorrect} answer, in this setting, a decrease in accuracy indicates that the judge is pulled toward position A and therefore toward the incorrect option. At the same time, an increase suggests the judge resists the bias or becomes more reliable despite it. Across the three tasks, bandwagon, distraction, diversity, and verbosity increase Overall accuracy compared to no-bias, whereas authority, CoT, final-only, refined, and sentiment decrease Overall accuracy. 

All three tasks still follow the pattern of having accuracy scores of easy $>$ medium $>$ hard without any bias. Bias effects are again most pronounced on hard instances. On CodeRepair hard, several biases sharply reduce accuracy, including refined with a $-17.80$ drop, CoT with a $-13.23$ drop, and authority with a $-11.82$ drop relative to the no-bias hard baseline. This is consistent with the idea that hard cases provide weaker evidence, so the judge relies more on the bias attached to option A. Since A is incorrect here, it more easily pulls decisions toward the wrong option. However, in TestGen hard, multiple biases substantially improve accuracy, with distraction and verbosity increasing hard accuracy to $75.27$ from $59.24$, and diversity and self-enhancement also provide clear gains, which is consistent with these biases being less seeking in the B-correct setting. They reduce attraction to the biased option A, so the judge more often selects the correct option B on hard instances.

When comparing tasks, CodeRepair remains the highest-accuracy task overall for no-bias, followed by CodeGen and TestGen. CodeRepair also exhibits the strongest negative sensitivity under several biases when the bias conflicts with the gold answer, with refined and CoT producing large Overall and hard-case drops. CodeGen shows a similar but milder pattern, with the most damaging effects again coming from refined and sentiment. TestGen displays the widest spread, often improving under distraction, diversity, and verbosity, but also suffering the single largest Overall drop under sentiment at $-10.74$. Overall, many biases exhibit a consistent reversal. The direction of their accuracy change flips when the correct answer moves from A to B (and vice versa). Yet the magnitude and even the direction of the effect remain bias- and task-dependent, showing that the biases have a significant effect, but in different ways.

\begin{table*}[t]
\centering
\setlength{\tabcolsep}{4pt}
\resizebox{\textwidth}{!}{
\begin{tabular}{lrrrrrrrrrrrr}
\toprule
\multirow{2}{*}{\textbf{Bias (row)}} &
\multicolumn{4}{c}{\textbf{CodeGen}} &
\multicolumn{4}{c}{\textbf{CodeRepair}} &
\multicolumn{4}{c}{\textbf{TestGen}} \\
\cmidrule(lr){2-5}\cmidrule(lr){6-9}\cmidrule(lr){10-13}
 & \textbf{Easy} & \textbf{Medium} & \textbf{Hard} & \textbf{Overall}
 & \textbf{Easy} & \textbf{Medium} & \textbf{Hard} & \textbf{Overall}
 & \textbf{Easy} & \textbf{Medium} & \textbf{Hard} & \textbf{Overall} \\
\midrule
no-bias     & 91.80 & 77.99 & 52.12 & 75.41 & 96.17 & 86.71 & 54.52 & 79.80 & 88.89 & 78.33 & 59.24 & 71.14 \\
Authority & 91.69 & 73.15 & 47.77 & 71.68\down & 93.40 & 84.67 & 42.70 & 73.44\down & 89.89 & 64.71 & 58.42 & 67.72\down \\
Bandwagon & 93.46 & 80.61 & 50.15 & 75.97\up & 95.57 & 90.04 & 64.89 & 84.29\up & 86.32 & 80.36 & 65.41 & 73.81\up \\
CoT      & 90.80 & 75.59 & 42.58 & 69.80\down & 94.12 & 79.23 & 41.29 & 71.40\down & 81.32 & 75.41 & 56.22 & 66.47\down \\
Distraction & 94.15 & 85.42 & 58.42 & 79.82\up & 94.39 & 81.29 & 56.52 & 80.88\up & 88.42 & 83.64 & 75.27 & 80.42\up \\
Diversity & 95.32 & 83.14 & 57.46 & 79.06\up & 96.68 & 86.32 & 59.12 & 81.22\up & 89.89 & 85.71 & 66.31 & 75.69\up \\
Final-only & 91.07 & 80.26 & 52.44 & 74.31\down & 94.91 & 85.02 & 54.41 & 78.30\down & 90.00 & 79.37 & 55.17 & 70.03\down \\
Model-name & 92.67 & 82.75 & 54.31 & 75.68\up & 95.40 & 87.36 & 50.93 & 78.69\down & 89.77 & 77.59 & 63.21 & 72.57\up \\
Refined   & 89.33 & 68.91 & 37.73 & 65.49\down & 93.98 & 73.98 & 36.72 & 67.54\down & 89.90 & 70.00 & 59.49 & 69.77\down \\
Self-enhance & 91.10 & 77.44 & 49.11 & 73.86\down & 94.63 & 89.47 & 52.66 & 79.19\down & 90.12 & 71.43 & 66.84 & 73.29\up \\
Sentiment & 90.42 & 68.60 & 40.94 & 66.85\down & 93.66 & 71.67 & 47.31 & 73.82\down & 76.70 & 56.14 & 52.69 & 60.40\down \\
Verbosity & 94.84 & 83.86 & 60.40 & 80.24\up & 94.29 & 88.44 & 60.18 & 81.34\up & 91.21 & 76.12 & 75.27 & 79.71\up \\
\bottomrule
\end{tabular}
}
\caption{Biases vs.\ Tasks (accuracy \%) when the right answer is located at position B. Each task reports Easy/Medium/Hard/Overall for this model. Note that the bias is always towards position A. So in this case, the model is biased toward the incorrect answer.}
\label{tab:bias-b-qwen3-4b}
\end{table*}

\paragraph{Comparing position A and B}
Above, we compared the swapped position settings where the correct or incorrect answers are placed in position A or B (position A denotes the first candidate shown to the judge, and position B denotes the second). A robust judge would exhibit similar accuracy when the correct answer is placed at A versus at B, whereas large gaps indicate sensitivity to positional or prompt-induced cues. Even under no-bias, accuracies are higher when the gold answer is at B, suggesting an inherent order effect in the baseline prompt. This observation is aligned with previous studies~\cite{ye2024, jiang2025codejudge}.

Several biases show a clear directionality in this swap position setting. CoT, refined, and sentiment increase accuracy most of the time when the gold answer is at A, but decrease it when the gold answer is at B. For example, their Overall gains under position A range from +5.21 to +9.60, while under position B, the Overall score drops ranging from $-4.67$ to $-12.26$ across the tasks. In contrast, bandwagon, distraction, diversity, and verbosity behave in the opposite direction, decreasing accuracy when the gold answer is at A but increasing it when the gold answer is at B. Their Overall deltas are negative under position A but turn positive under position B, with the most pronounced gains on TestGen, where distraction and verbosity improve Overall accuracy. 
This reversal indicates that these biases suppress the judge's baseline preference for option A. Consequently, when A is correct, accuracy is reduced, whereas when B is correct, accuracy is enhanced.
Final-only, model-name, and self-enhance show smaller and more mixed A--B differences, indicating weaker and less consistent positional coupling than the strongly directional biases above.

Overall, the A versus B comparison shows that the same bias can either increase or decrease the accuracy depending on where the gold answer is placed in the prompt, and the magnitude of these sign flips provides direct evidence of position sensitivity and consistent preferences for these biases. 
Although the position of the candidate is a fundamental carrier in this pairwise prompting setting, the bias conditions are content-level interventions attached to a position, and their effects are evident even when the correct position is held fixed. Within the same position setting, different biases can change the accuracy by various magnitudes across tasks. This change in accuracy indicates that the judge is sensitive to multiple non-position biases, such as CoT and bandwagon.

\begin{tcolorbox}[colback=gray!10,colframe=gray!50,boxrule=0.6pt,arc=2pt,
                  left=6pt,right=6pt,top=6pt,bottom=6pt]
\noindent\textbf{Key findings:}
Qwen3-4B exhibits pronounced positional sensitivity, with judgment difficulty systematically decreasing when the correct answer resides in position B rather than A, even in the absence of bias injection. Beyond this sequential effect, we observe significant effect of injected bias, with several injected biases exhibiting opposite directions in B-correct versus A-correct scenarios. This indicates that these bias has a  consistent effects on the judge rather than uniform performance modulators.
\end{tcolorbox}
\subsubsection{Qwen2.5-Coder-3B}
In this section, we will discuss the obtained results for Qwen2.5-Coder-3B, i) when the correct answer is at position A, ii) correct answer is placed at position B, and iii) comparing the results.

\begin{table*}[t]
\centering
\setlength{\tabcolsep}{4pt}
\resizebox{\textwidth}{!}{
\begin{tabular}{lrrrrrrrrrrrr}
\toprule
\multirow{2}{*}{\textbf{Bias (row)}} &
\multicolumn{4}{c}{\textbf{CodeGen}} &
\multicolumn{4}{c}{\textbf{CodeRepair}} &
\multicolumn{4}{c}{\textbf{TestGen}} \\
\cmidrule(lr){2-5}\cmidrule(lr){6-9}\cmidrule(lr){10-13}
 & \textbf{Easy} & \textbf{Medium} & \textbf{Hard} & \textbf{Overall}
 & \textbf{Easy} & \textbf{Medium} & \textbf{Hard} & \textbf{Overall}
 & \textbf{Easy} & \textbf{Medium} & \textbf{Hard} & \textbf{Overall} \\
\midrule
no-bias      & 62.96 & 64.29 & 57.00 & 60.99 & 75.14 & 66.19 & 62.30 & 67.38 & 60.66 & 57.79 & 64.54 & 62.41 \\
Authority   & 82.34 & 79.58 & 73.56 & 78.12\up & 85.77 & 83.26 & 80.08 & 82.76\up & 87.50 & 80.12 & 80.29 & 81.85\up \\
Bandwagon    & 62.39 & 59.58 & 53.90 & 58.27\down & 73.53 & 66.29 & 59.41 & 65.74\down & 48.91 & 49.06 & 54.49 & 52.19\down \\
CoT          & 78.90 & 74.39 & 69.12 & 73.80\up & 83.85 & 78.46 & 75.73 & 79.02\up & 74.86 & 79.01 & 73.78 & 75.03\up \\
Distraction  & 58.81 & 56.60 & 49.88 & 54.67\down & 69.40 & 63.18 & 62.75 & 64.91\down & 50.83 & 46.58 & 47.42 & 48.00\down \\
Diversity    & 54.25 & 48.06 & 43.08 & 48.14\down & 67.95 & 58.21 & 53.45 & 59.29\down & 41.21 & 35.22 & 40.98 & 39.93\down \\
Final-only   & 55.01 & 48.35 & 44.66 & 49.09\down & 68.95 & 55.62 & 55.28 & 59.54\down & 53.55 & 48.43 & 49.48 & 50.18\down \\
Model-name   & 69.74 & 66.49 & 62.82 & 66.12\up & 78.52 & 72.39 & 67.95 & 72.49\up & 65.93 & 51.61 & 61.57 & 60.66\down \\
Refined      & 89.26 & 88.35 & 84.00 & 86.93\up & 90.77 & 89.32 & 84.13 & 87.68\up & 86.26 & 89.24 & 88.45 & 88.12\up \\
Self-enhance & 68.60 & 62.37 & 60.10 & 63.54\up & 74.45 & 66.62 & 62.51 & 67.38 & 53.98 & 63.12 & 55.94 & 56.92\down \\
Sentiment    & 95.38 & 92.19 & 90.45 & 92.56\up & 95.25 & 92.85 & 91.30 & 92.96\up & 90.11 & 85.35 & 87.96 & 87.94\up \\
Verbosity    & 30.64 & 30.57 & 29.99 & 30.36\down & 56.45 & 49.57 & 48.34 & 51.19\down & 31.52 & 23.60 & 23.63 & 25.36\down \\
\bottomrule
\end{tabular}
}
\caption{Bias rows vs.\ tasks (accuracy \%) on examples with right answer is at position A for Qwen2.5-Coder-3B. }
\label{tab:bias-coder-A}
\end{table*}

\paragraph{A correct}
For Qwen2.5-Coder-3B, across the three tasks in ~\autoref{tab:bias-coder-A}, four biases consistently increase while five consistently decrease the accuracy. At the Overall level, Authority, CoT, Refined, and Sentiment yield uniformly positive deltas across CodeGen, CodeRepair, and TestGen, suggesting that these biases tend to be favored by the judge so that the prediction is more aligned with the gold position under this setting. In contrast, verbosity exhibits the strongest negative shifts, with bandwagon, distraction, diversity, and final-only also consistently decreasing accuracy, indicating that these biases tend to pull the judge away from the correct answer. Meanwhile, model-name and self-enhance are not uniformly directional, they are mildly positive on some tasks but become neutral or negative on others, and are therefore better characterized as task-dependent.

Across difficulty strata, bias effects are typically more pronounced on harder instances. In particular, the A-aligned bias remain positive on hard subsets, whereas the adverse bias can substantially deflate hard accuracy. Finally, at the task level, CodeRepair has the highest no-bias Overall baseline (67.38).
It also shows fewer extreme collapses under bias, with the highest worst-case Overall accuracy (51.19 under verbosity).
In contrast, TestGen shows the widest spread across biases (Overall: 25.36--88.12), making it the most cue-sensitive setting in this A-correct evaluation.

\begin{table*}[t]
\centering
\setlength{\tabcolsep}{4pt}
\resizebox{\textwidth}{!}{
\begin{tabular}{lrrrrrrrrrrrr}
\toprule
\multirow{2}{*}{\textbf{Bias (row)}} &
\multicolumn{4}{c}{\textbf{CodeGen}} &
\multicolumn{4}{c}{\textbf{CodeRepair}} &
\multicolumn{4}{c}{\textbf{TestGen}} \\
\cmidrule(lr){2-5}\cmidrule(lr){6-9}\cmidrule(lr){10-13}
 & \textbf{Easy} & \textbf{Medium} & \textbf{Hard} & \textbf{Overall}
 & \textbf{Easy} & \textbf{Medium} & \textbf{Hard} & \textbf{Overall}
 & \textbf{Easy} & \textbf{Medium} & \textbf{Hard} & \textbf{Overall} \\
\midrule
no-bias      & 51.24 & 46.41 & 36.87 & 44.27 & 60.41 & 51.22 & 39.31 & 49.27 & 41.53 & 32.28 & 38.59 & 38.03 \\
Authority    & 27.78 & 23.83 & 20.61 & 23.86\down & 34.97 & 26.79 & 19.88 & 26.52\down & 19.34 & 24.38 & 19.14 & 20.19\down \\
Bandwagon   & 50.07 & 50.96 & 42.18 & 47.22\up & 55.30 & 47.71 & 37.83 & 46.09\down & 54.64 & 48.77 & 50.83 & 51.27\up \\
CoT          & 34.39 & 33.04 & 28.38 & 31.65\down & 45.64 & 32.01 & 23.21 & 32.64\down & 23.50 & 24.69 & 23.98 & 24.01\down \\
Distraction  & 54.05 & 49.65 & 48.04 & 50.48\up & 61.28 & 52.65 & 38.44 & 49.60\up & 58.89 & 59.01 & 54.53 & 56.35\up \\
Diversity    & 63.53 & 57.91 & 53.42 & 58.01\up & 65.08 & 56.22 & 45.65 & 54.73\up & 65.93 & 55.56 & 58.16 & 59.35\up \\
Final-only   & 61.25 & 57.69 & 51.03 & 56.24\up & 64.35 & 58.71 & 46.79 & 55.64\up & 51.63 & 50.31 & 51.44 & 51.26\up \\
Model-name   & 44.38 & 39.90 & 33.09 & 38.70\down & 54.97 & 43.41 & 32.10 & 42.48\down & 39.01 & 36.31 & 36.42 & 36.97\down \\
Refined      & 19.30 & 16.35 & 15.01 & 16.79\down & 29.85 & 21.00 & 11.71 & 19.98\down & 14.36 & 10.62 & 13.96 & 13.41\down \\
Self-enhance & 52.55 & 50.43 & 38.00 & 46.23\up & 56.07 & 46.99 & 38.20 & 46.25\down & 37.16 & 44.59 & 39.63 & 40.02\up \\
Sentiment    & 9.68  & 7.77  & 7.87  & 8.44\down  & 17.93 & 13.98 & 7.67  & 12.66\down & 12.57 & 11.18 & 12.45 & 12.23\down \\
Verbosity    & 76.05 & 74.61 & 66.95 & 72.07\up & 72.46 & 66.00 & 50.41 & 61.72\up & 74.73 & 79.63 & 72.87 & 74.58\up \\
\bottomrule
\end{tabular}
}
\caption{Bias rows vs.\ tasks (accuracy \%) on examples with right answer is at position B for Qwen2.5-Coder-3B.}
\label{tab:bias-coder-B}
\end{table*}

\paragraph{B correct}
Table~\ref{tab:bias-coder-B} reveals a pronounced directional reversal relative to the A-correct setting, indicating strong Bias$\times$Position interactions. When the gold answer is at B, bias that tend to help the judge align with the gold position under A-correct become consistently adverse: Authority, CoT, refined, and sentiment and often model-name (except for TestGen medium) produce uniformly negative shifts across CodeGen, CodeRepair, and TestGen. In other words, when these bias are attached to option A, they systematically pull the judge away from the correct B. 

Conversely, verbosity shows the most consistent positive shifts across all three tasks, and distraction, diversity, and final-only exhibit the same cross-task pattern.
Since our bias cue is always attached to position A, this consistency suggests that these prompts act as negative cues for Qwen2.5-Coder-3B in the B-correct condition.
They reduce the perceived credibility of option A, which counteracts the judge's no-bias tendency to select A, and thus makes the judge more likely to choose B (the correct option in this setting). In contrast, bandwagon and self-enhance are less stable, as they follow this direction on some tasks but not others, indicating that the same cue can be interpreted differently across tasks and difficulty strata.
In Table~\ref{tab:bias-coder-B}, the no-bias baseline is below 50 on all three tasks (Overall: 38.03--49.27).
For a balanced A/B choice, 50 is the chance level.
This indicates a strong order effect for Qwen2.5-Coder-3B: under uncertainty, the judge tends to select option A more often than option B.
When the gold answer is placed at B, this A-preference directly translates into below-chance accuracy.

A plausible explanation is that these prompt interventions are not neutral features whose effects are independent of position. Rather, they operate through how the judge interprets the presentation of the candidate's answer at A. Bias, such as CoT or refined, may increase the perceived completeness, confidence, or legitimacy of the response, which helps when A is gold but becomes harmful when B is gold. In contrast, verbosity or distraction may be perceived as over-elaboration, noise, uncertainty, or a style mismatch, which makes the response look less trustworthy and thus suppresses A-preference. This suppression can hurt under A-correct but becomes beneficial under B-correct by shifting selections toward B. Overall, TestGen shows one of the widest cross-bias spreads, especially on hard instances, reinforcing it as a bias sensitive setting and motivating position-stratified reporting.
\paragraph{Comparing position A and B.}
\autoref{tab:bias-coder-A} and \autoref{tab:bias-coder-B} contrast the swapped-position evaluation where the gold answer is placed at position A (first candidate) versus at position B (second candidate). A position-robust judge would show similar accuracy under these two settings, whereas large gaps indicate that the judgment is coupled to presentation cues that correlate with the displayed order.

For Qwen2.5-Coder-3B, the no-bias baseline already exhibits a strong order effect, in the opposite direction of Qwen3-4B, where A is easier. When the gold answer is at position A, accuracy is consistently higher across CodeGen, CodeRepair, and TestGen than when the gold answer is at position B. This implies an inherent preference toward selecting the first candidate under the baseline prompt, which makes correctness substantially easier to achieve when the better response is shown as A, and substantially harder when the better response is shown as B.

Several bias conditions amplify this baseline A preference in a highly directional way. Authority, CoT, refined, and sentiment align much more closely with the gold outcome when the gold answer is at A, reaching very high overall accuracies across all three tasks in \autoref{tab:bias-coder-A}. Under the swapped setting where the gold answer is at B, the same interventions sharply deviate from gold alignment in \autoref{tab:bias-coder-B}, often collapsing to far below the no-bias baseline. Refined and sentiment are the most extreme cases, with near-ceiling performance when the correct answer is presented as A, but near-floor performance when the correct answer is presented as B. This sign reversal indicates that these interventions do not simply improve judging quality, but rather strengthen a consistent preference for the first position, which helps only when the gold happens to be placed at A.

In contrast, another group of biases counteracts the baseline preference for position A and therefore shows the opposite directionality. Verbosity, diversity, and final-only strongly reduce accuracy when the gold answer is at A in \autoref{tab:bias-coder-A}, yet they become reliably more gold-aligned when the gold answer is at B in \autoref{tab:bias-coder-B}. Verbosity is particularly diagnostic because it is among the worst settings when the correct answer is A, but among the best when the correct answer is B, which is consistent with it suppressing the model's tendency to choose the first candidate. Distraction and bandwagon show a similar but more task-dependent pattern, with clearer benefits under position B on TestGen than on the other tasks. Model-name and self-enhance remain comparatively mixed and smaller in magnitude, suggesting weaker and less consistent coupling to the A versus B placement than the strongly directional groups above.

Overall, the results show large and systematic sign flips between \autoref{tab:bias-coder-A} and \autoref{tab:bias-coder-B}. These flips provide direct evidence that Qwen2.5-Coder-3B is position-sensitive, and that content-level prompt interventions can either reinforce or counteract an existing positional preference. Importantly, the magnitude of the shifts within a fixed gold position demonstrates that the observed effects cannot be attributed to position alone, because different biases materially change gold alignment even when the correct position is held fixed.

\begin{tcolorbox}[colback=gray!10,colframe=gray!50,boxrule=0.6pt,arc=2pt,
                  left=6pt,right=6pt,top=6pt,bottom=6pt]
\noindent\textbf{Key findings: }
Qwen2.5-Coder-3B is \textbf{highly position-dependent}: the no-bias baseline drops sharply from A-correct to B-correct.
When \textbf{A} is correct, refined/sentiment/authority/coT strongly boost all tasks,
but when \textbf{B} is correct, these same cues \textbf{collapse} accuracy,
while verbosity (and to a lesser extent Diversity/Final-only/Distraction) \textbf{reverses the trend} and dominates on B-correct.
\end{tcolorbox}

\subsubsection{GPT}

\begin{table*}[t]
\centering
\setlength{\tabcolsep}{4pt}
\resizebox{\textwidth}{!}{
\begin{tabular}{lrrrrrrrrrrrr}
\toprule
\multirow{2}{*}{\textbf{Bias (row)}} &
\multicolumn{4}{c}{\textbf{CodeGen}} &
\multicolumn{4}{c}{\textbf{CodeRepair}} &
\multicolumn{4}{c}{\textbf{TestGen}} \\
\cmidrule(lr){2-5}\cmidrule(lr){6-9}\cmidrule(lr){10-13}
 & \textbf{Easy} & \textbf{Medium} & \textbf{Hard} & \textbf{Overall} 
 & \textbf{Easy} & \textbf{Medium} & \textbf{Hard} & \textbf{Overall}
 & \textbf{Easy} & \textbf{Medium} & \textbf{Hard} & \textbf{Overall} \\
\midrule
no-bias         & 96.85 & 88.00 & 60.24 & 79.98 & 97.32 & 91.39 & 65.82 & 82.93 & 95.83 & 87.81 & 67.23 & 77.46 \\
Authority   & 97.50 & 95.83 & 61.23 & 84.16\up & 97.50 & 94.84 & 72.38 & 87.72\up & 87.75 & 81.15 & 59.62 & 69.32\down \\
Bandwagon   & 93.75 & 81.63 & 50.46 & 73.78\down & 97.23 & 84.17 & 59.87 & 78.32\down & 82.21 & 66.38 & 56.50 & 62.83\down \\
CoT        & 97.50 & 96.89 & 65.37 & 86.91\up & 97.50 & 95.09 & 83.82 & 91.36\up & 97.50 & 89.14 & 73.23 & 82.67\up \\
Distraction (A) & 93.37 & 88.91 & 53.41 & 76.46\down & 95.82 & 82.67 & 57.55 & 76.79\down & 88.32 & 62.50 & 48.72 & 60.96\down \\
Diversity & 95.74 & 89.91 & 52.51 & 77.85\down & 96.07 & 85.91 & 63.79 & 79.90\down & 90.30 & 68.62 & 56.71 & 66.16\down \\
Final-only  & 96.24 & 92.81 & 57.60 & 80.95\up & 97.50 & 86.41 & 67.32 & 82.62\down & 97.50 & 85.59 & 60.80 & 74.97\down \\
Model-name  & 96.79 & 94.36 & 60.63 & 82.51\up & 97.50 & 91.01 & 71.59 & 86.27\up & 87.90 & 69.28 & 51.83 & 62.51\down \\
Refined     & 97.50 & 95.66 & 74.97 & 89.20\up & 97.50 & 97.30 & 84.40 & 92.53\up & 97.50 & 82.25 & 68.40 & 76.83\down \\
Self-enhance& 96.57 & 91.77 & 56.69 & 79.92\down & 97.50 & 88.63 & 63.83 & 82.10\down & 97.47 & 76.35 & 66.34 & 74.52\down \\
Sentiment   & 97.50 & 97.50 & 70.50 & 88.86\up & 97.50 & 95.04 & 83.22 & 92.10\up & 97.50 & 83.20 & 77.61 & 84.35\up \\
Verbosity   & 89.08 & 83.08 & 45.75 & 70.04\down & 96.77 & 81.68 & 59.70 & 77.68\down & 92.76 & 74.31 & 47.27 & 61.72\down \\
\bottomrule
\end{tabular}
}
\caption{GPT LLM-as-a-Judge under different prompt biases.}
\label{tab:bias_gpt}
\end{table*}

For GPT-based judges, the neutral prompt already yields strong accuracy, but the scores vary substantially with both task difficulty and prompt-level presentation cues. Under no-bias, Overall accuracy is around 80 across tasks (CodeGen 79.98, CodeRepair 82.93, TestGen 77.46). Across the results, we can observe that difficulty is the dominant factor. Easy subsets are close to saturation (95 to 97), while Hard subsets are much lower (60.24 on CodeGen, 65.82 on CodeRepair, 67.23 on TestGen). The Easy-to-Hard gap is largest for CodeGen (96.85 to 60.24), indicating that code generation judgments are the most difficulty-sensitive among the three tasks, which aligns with a previous study~\cite{jiang2025codejudge}.

Across biases, task and hardness interact strongly. Prompts that inject explicit reasoning structure or self-checking tend to raise accuracy most clearly on Hard items, especially for CodeGen and CodeRepair. CoT increases the hard set on CodeRepair from 65.82 to 83.82 and also lifts TestGen's hard set to 73.23. Refined attains the highest CodeGen hard score (74.97) and raises CodeRepair's overall accuracy to 92.53. Sentiment is the strongest condition for TestGen, increasing TestGen hard score to 77.61 and TestGen's overall score to 84.35. These improvements are not uniform across difficulty, and several settings mainly benefit Hard subsets while leaving Easy nearly unchanged because it is already saturated.

A second group of biases shifts behavior more mildly and often in a task-dependent way.
Authority cues and Model-name exposure slightly increase Overall accuracy for CodeGen and CodeRepair, but they substantially reduce it for TestGen.
For example, Model-name lifts CodeRepair Overall from 82.93 to 86.27, yet TestGen Overall drops from 77.46 to 62.51 ($-14.95$).
Final-only and Self-enhance stay closer to the no-bias baseline on CodeGen/CodeRepair, but they still reduce TestGen accuracy.
In general, these cues proviguidance at the presentation-level rather than better task reasoning.
They can change which option looks more credible, and this shift does not transfer reliably across tasks.

For GPT, the neutral prompt already yields strong accuracy across tasks, but the judge remains sensitive to prompt interventions, which can noticeably shift its decisions. Prompts that add explicit reasoning structure often push accuracy upward, while authority- or model-identity cues can pull it downward, so these changes should be read as evidence of bias susceptibility rather than unqualified improvement.

\begin{tcolorbox}[colback=gray!10,colframe=gray!50,boxrule=0.6pt,arc=2pt,
                  left=6pt,right=6pt,top=6pt,bottom=6pt]
\noindent\textbf{Key findings (GPT): }
GPT shows a \textbf{strong no-bias baseline} across tasks, with \textbf{Easy} nearly saturated while \textbf{Hard} remains substantially lower.
Biases that inject explicit reasoning structure yield the highest measured accuracies on CodeGen and CodeRepair, and most noticeably lift performance on \textbf{Hard} subsets.
In contrast, TestGen is the most fragile: several biases produce large drops in Overall accuracy, with \textbf{verbosity} showing the sharpest decline.
\end{tcolorbox}


\subsection{Effect of Bias on LLM Judgment, Analysis Across Tasks}
\begin{figure*}[t]
  \centering
  \includegraphics[width=\textwidth]{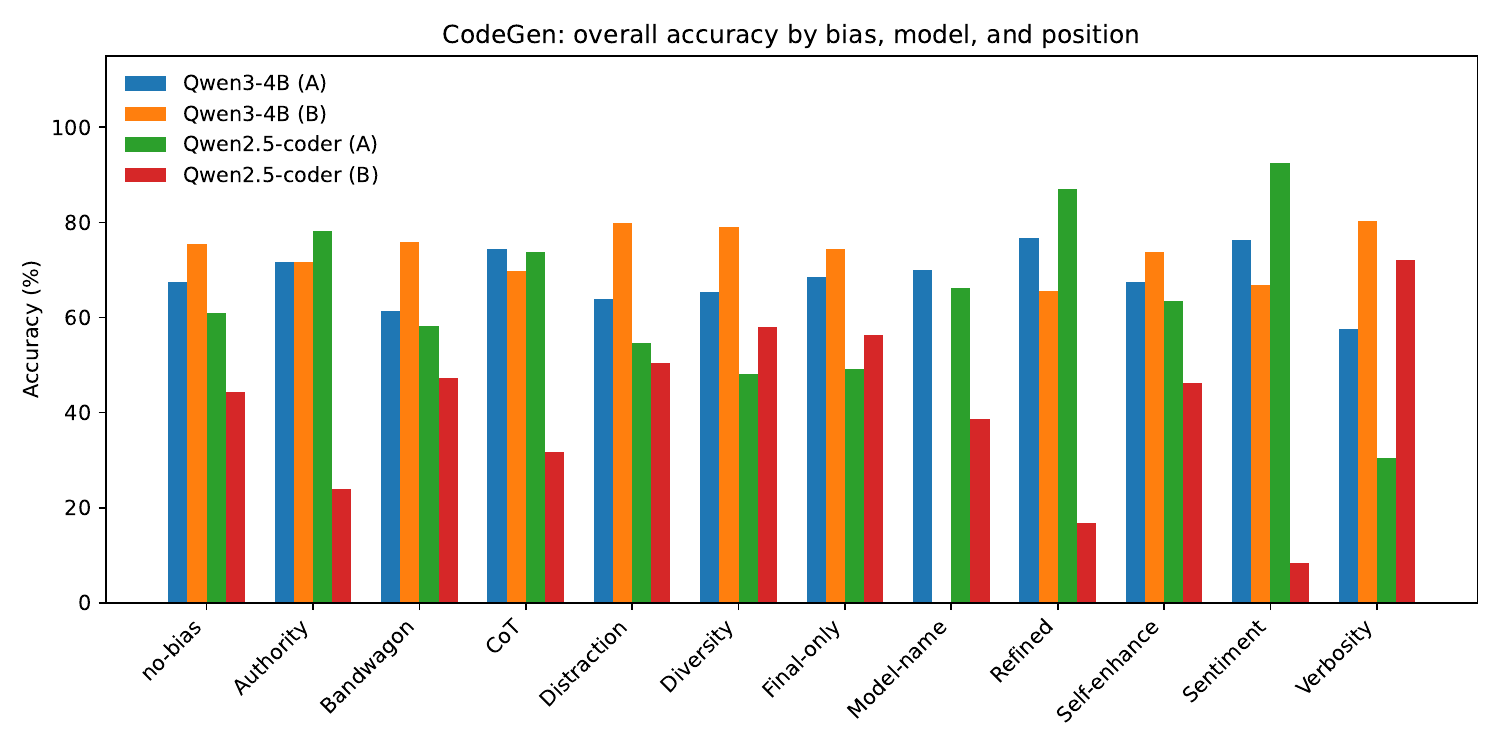}
  \caption{CodeGen: Overall accuracy by bias, model, and position (A vs.\ B).}
  \label{fig:bias-by-task-codegen}
\end{figure*}

For CodeGen (Figure \ref{fig:bias-by-task-codegen}), the dominant pattern is that both judges become highly asymmetric under several biases. When the correct answer is at position A, CoT, Authority, Refined, and especially Sentiment produce very high accuracies, whereas the same biases push performance on B-labelled examples down to very low levels for Qwen2.5-Coder-3B. In contrast, Verbosity shows the opposite shape: both models are poor on A-labelled examples but suddenly become strong on B-labelled examples, indicating that verbosity cues systematically steer the judge toward B. Qwen2.5-Coder-3B exhibits the same qualitative trends as Qwen3-4B, but the bars are more extreme: its Sentiment and Refined biases almost saturate on A while collapsing on B, and its Verbosity bias produces the largest B-favouring effect. Overall, CodeGen already shows that the direction of the injected cue (A vs.\ B) matters far more than the underlying model version; the two judges share the same failure modes, with Qwen2.5-Coder-3B simply amplifying them.

\begin{figure*}[t]
  \centering
  \includegraphics[width=\textwidth]{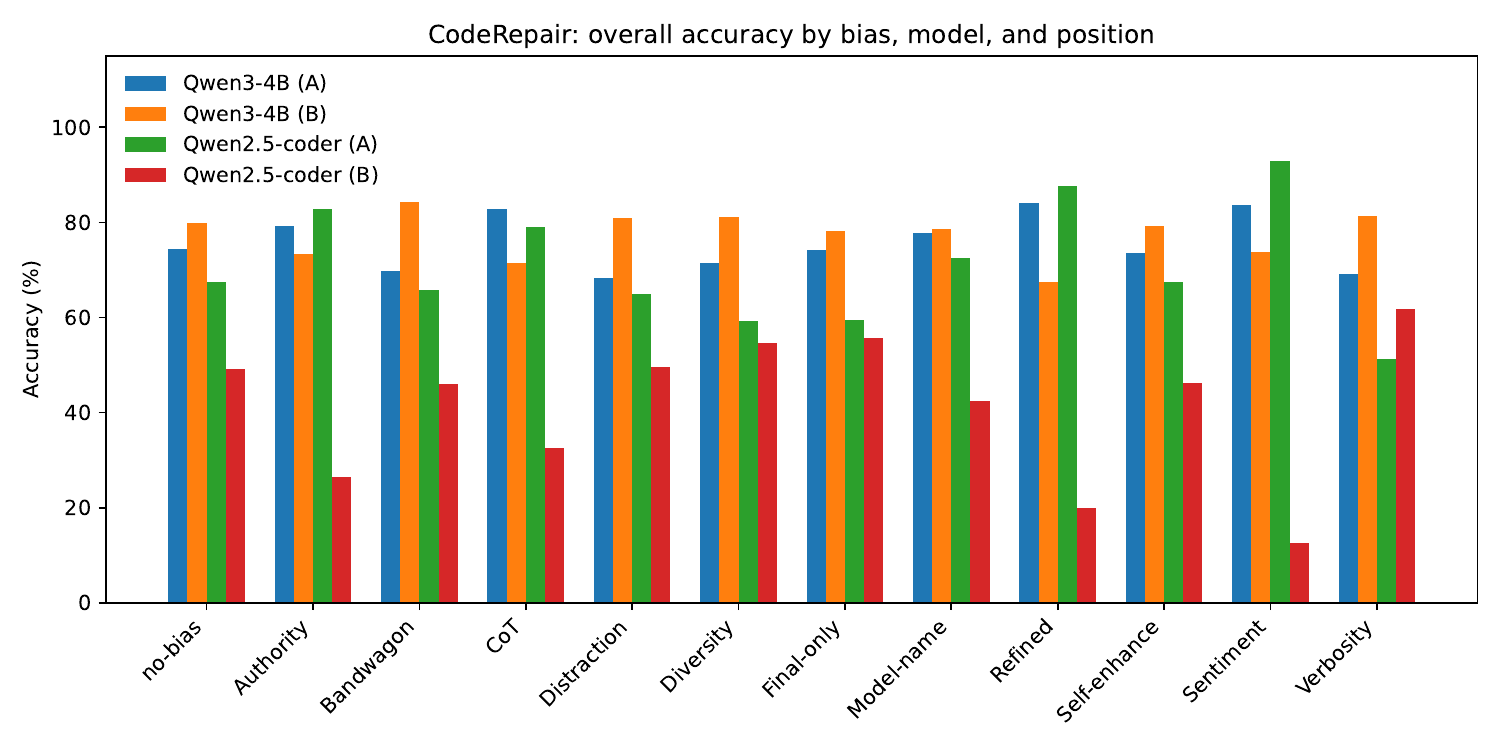}
  \caption{CodeRepair: Overall accuracy by bias, model, and position (A vs.\ B).}
  \label{fig:bias-by-task-coderepair}
\end{figure*}

On CodeRepair (Figure \ref{fig:bias-by-task-coderepair}), the absolute accuracies are higher across the board, but the cross-model, cross-position structure remains similar. Both models achieve strong performance on A-labelled examples under CoT, Authority, Refined, and Sentiment, and again perform very poorly on B-labelled examples in exactly those rows. Verbosity continues to favour B, although the gap between A and B is slightly smaller than in CodeGen. Diversity and Final-only appear comparatively benign: they move A and B in the same direction and maintain a smaller separation, suggesting that they encode a milder preference signal. The key observation is that, on this easier task, biases do not merely add noise but induce consistent, systematic shifts in which side the judge prefers; those shifts are largely shared between Qwen3 and Qwen2.5, which again differ mainly in magnitude.

\begin{figure*}[t]
  \centering
  \includegraphics[width=\textwidth]{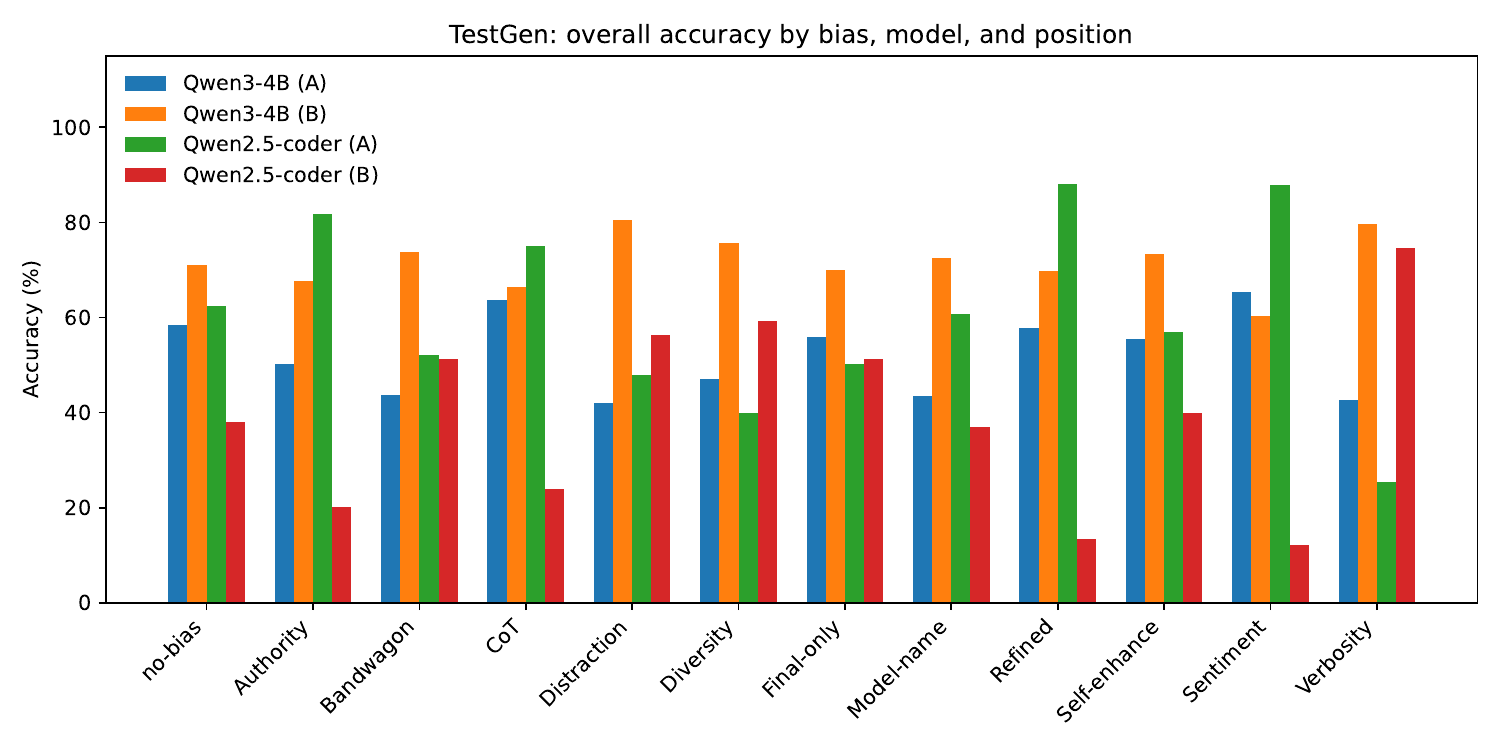}
  \caption{TestGen: Overall accuracy by bias, model, and position (A vs.\ B).}
  \label{fig:bias-by-task-testgen}
\end{figure*}

The TestGen plot (Figure \ref{fig:bias-by-task-testgen}) shows the strongest sensitivity to injected biases. Baseline performance is lower than for CodeRepair, and the relative spread across biases is larger. As in the other tasks, Refined and Sentiment produce near-ceiling performance on A and near-floor performance on B for both models, while CoT and Authority also show large A–B gaps. Verbosity is particularly striking here: for both judges, accuracy on B-labelled examples under Verbosity is among the highest of any condition, while accuracy on A-labelled examples is among the lowest, effectively flipping the judge’s preference. Bandwagon, Distraction, and Diversity sit in the middle; they still create visible gaps between A and B but with more moderate strength. Taken together, the three plots show that bias effects are highly consistent across Qwen3-4B and Qwen2.5-Coder-3B and across tasks, but their impact scales with task difficulty: CodeRepair is relatively robust, CodeGen is moderately exposed, and TestGen is the most vulnerable to systematic prompt-induced shifts in which the judge favours.

\paragraph{Cross-task patterns of bias.}
Looking across the three tasks in Figures~\ref{fig:bias-by-task-codegen}--\ref{fig:bias-by-task-testgen}, the most striking phenomenon is how stable the bias ordering is across settings. For both Qwen3-4B and Qwen2.5-Coder-3B, the same subset of prompts---CoT, Authority, Refined, and Sentiment---consistently produce very high accuracy when the correct answer is at position~A and very low accuracy when the correct answer is at position~B, regardless of whether the underlying task is CodeGen, CodeRepair, or TestGen. In other words, once the judge is told that system~A is authoritative, refined, or emotionally preferred, it tends to choose~A on \emph{all} instances, so these cues help on A-labelled examples and systematically hurt on B-labelled ones in every task.

The complementary pattern is equally consistent. Verbosity behaves as a robust B-favouring bias across all three tasks: for both models, the bars for B-labelled examples under Verbosity are always among the highest in the plot, while the corresponding A-labelled bars are among the lowest. Prompts that reward longer responses, therefore flip the judge's preference toward the second position in a way that is not task-specific. Milder prompts such as Bandwagon, Distraction, Diversity, Final-only, Model-name, and Self-enhance show the same qualitative direction of effect across tasks and models; for example, Diversity and Final-only slightly favour B over A in CodeGen, CodeRepair, and TestGen, while Bandwagon and Distraction create smaller but still visible A/B gaps, but with a smaller magnitude than the strongly polarising biases above.

These patterns across tasks suggest that the injected meta-information makes the judge generally trust one position more, rather than affecting only a specific benchmark. CodeRepair has higher absolute accuracies, and TestGen is more fragile, yet the relative ordering of biases and the sign of their A-versus-B gaps are essentially preserved. The same bias that makes Qwen3-4B over-trust position~A on CodeGen also makes Qwen2.5-Coder-3B over-trust position~A on TestGen; similarly, the same verbosity cue that rescues B-labelled examples on TestGen also lifts B on CodeGen and CodeRepair. This consistency across tasks and model variants indicates that prompt-induced biases are a stable property of the judge configuration itself, rather than an artefact of a single dataset or task.

\begin{tcolorbox}[colback=gray!10,colframe=gray!50,boxrule=0.6pt,arc=2pt,
                  left=6pt,right=6pt,top=6pt,bottom=6pt]
\noindent\textbf{Key finding: }
Across all tasks, prompt biases act as a \emph{positional prior}: \textbf{CoT/Authority/Refined/Sentiment} strongly pushes the judge to choose \textbf{A} (near-ceiling on A-correct but near-floor on B-correct), while \textbf{Verbosity} consistently pushes toward \textbf{B}.  
These patterns are consistent across Qwen3-4B and Qwen2.5-Coder-3B, and the effect size grows with task fragility (TestGen $>$ CodeGen $>$ CodeRepair).
\end{tcolorbox}

\subsection{Consistent Results in LLM Judgment (RQ2 Results)}

In this section, we will report the results for RQ2.
\begin{table}[t]
\centering
\small
\setlength{\tabcolsep}{4pt}
\begin{tabular}{lcccccccccccc}
\toprule
\textbf{Bias (row)} & \multicolumn{4}{c}{\textbf{CodeGen (CR \%)}} & \multicolumn{4}{c}{\textbf{CodeRepair (CR \%)}} & \multicolumn{4}{c}{\textbf{TestGen (CR \%)}} \\
\cmidrule(lr){2-5}\cmidrule(lr){6-9}\cmidrule(lr){10-13}
& Easy & Med & Hard & Overall & Easy & Med & Hard & Overall & Easy & Med & Hard & Overall \\
\midrule
no-bias & 91.67 & 77.34 & 76.83 & 83.60 & 97.03 & 91.90 & 83.07 & 91.41 & 86.21 & 63.79 & 58.60 & 70.56 \\
Authority & 93.50 & 77.25 & 73.55 & 83.47 & 95.25 & 92.79 & 87.84 & 92.45 & 85.59 & 65.62 & 65.00 & 71.82 \\
Bandwagon & 93.63 & 73.60 & 75.29 & 83.25 & 96.38 & 93.33 & 83.85 & 91.81 & 87.10 & 65.96 & 68.33 & 74.64 \\
CoT & 91.69 & 80.56 & 74.06 & 83.41 & 96.10 & 91.34 & 88.25 & 92.50 & 87.10 & 67.21 & 66.67 & 74.06 \\
Distraction & 91.67 & 81.13 & 74.15 & 83.52 & 95.62 & 90.79 & 87.28 & 90.86 & 83.58 & 71.67 & 79.78 & 79.84 \\
Diversity & 92.26 & 81.41 & 71.56 & 83.31 & 95.82 & 89.76 & 86.21 & 91.34 & 90.68 & 74.42 & 71.59 & 78.64 \\
Final-only & 90.75 & 76.83 & 75.54 & 82.55 & 97.23 & 89.26 & 84.57 & 91.31 & 91.37 & 73.44 & 69.23 & 78.23 \\
Model-name & 91.09 & 77.07 & 71.56 & 81.46 & 96.03 & 90.86 & 85.79 & 91.74 & 80.67 & 65.52 & 65.05 & 70.25 \\
Refined & 91.38 & 82.59 & 72.28 & 83.14 & 97.42 & 91.39 & 85.94 & 92.36 & 85.95 & 69.12 & 63.89 & 72.09 \\
Self-enhance & 91.40 & 79.32 & 71.53 & 82.39 & 95.54 & 91.99 & 82.70 & 90.85 & 84.30 & 65.45 & 64.77 & 71.59 \\
Sentiment & 93.04 & 77.97 & 73.92 & 83.56 & 96.10 & 90.75 & 84.16 & 91.20 & 85.82 & 69.84 & 59.54 & 70.81 \\
Verbosity & 92.53 & 77.10 & 77.93 & 84.14 & 94.58 & 90.96 & 85.60 & 90.93 & 89.84 & 66.10 & 74.42 & 78.55 \\
\bottomrule
\end{tabular}
\caption{Consistency scores (CR, \%) across two independent runs under identical settings for Qwen3, reported by task and difficulty.}
\label{tab:consistency_cr_qwen}
\end{table}

\subsubsection{Qwen3-4B}
Across Table~\ref{tab:consistency_cr_qwen}, the judge exhibits generally high repeatability. Overall Consistency Rates (CR) typically land in the low-to-mid 80s for CodeGen, the low 90s for CodeRepair, and the low-to-high 70s for TestGen. This indicates that the model’s pairwise preferences are often reproducible across independent runs under identical settings. However, stability is clearly task-dependent. CodeRepair is the most stable across nearly all bias rows. Conversely, TestGen is consistently less stable, especially at the harder strata. This suggests that tasks with greater ambiguity in evaluation signals are more sensitive to sampling variability and prompt cues.

At the aggregate level, most biases in CodeGen and CodeRepair stay close to the no-bias baseline instead of causing large destabilization. For instance, bias like CoT, Refined, and Authority remain broadly comparable to the baseline across these two tasks. However, TestGen reveals a paradoxical trend. Several biases actually show a noticeable increase in stability compared to the baseline (70.56). Specifically, distraction (79.84), verbosity (78.55), final-only (78.23) and Diversity (78.64) yield higher CR. When the core evaluation signal is inherently ambiguous, the model may reliably anchor onto superficial formatting or linguistic cues to make its decision. Importantly, this highlights a critical distinction, that higher CR reflects repeatability, not correctness. A consistently repeated bias can still be systematically wrong. Meanwhile, a lower CR simply signals that single-run judgments are fluctuating closer to the decision boundary.

The clearest structural pattern aligns with task difficulty. For almost every bias row and task, CR peaks on Easy and decreases across Medium and Hard. With few exceptions, this downward trend is expected in pairwise judging. Hard cases tend to be closer calls. Consequently, minor shifts in attention or sampling can easily flip the model's preference. This decline is mild for CodeRepair, where even hard items remain relatively stable. In contrast, the drop is much sharper for TestGen. Here, hard items frequently fall into the high 50s to mid 70s, depending on the injected bias. This points to a stronger uncertainty regime in TestGen, where biases interact with borderline comparisons to amplify run-to-run variability. Ultimately, the table highlights a key reliability consideration: stability depends jointly on task difficulty and the specific prompt cues. In regimes with lower CR, reporting repeated-run statistics alongside accuracy becomes crucial.

\begin{table}[t]
\centering
\small
\setlength{\tabcolsep}{4pt}
\begin{tabular}{lcccccccccccc}
\toprule
\textbf{Bias (row)} & \multicolumn{4}{c}{\textbf{CodeGen (CR \%)}} & \multicolumn{4}{c}{\textbf{CodeRepair (CR \%)}} & \multicolumn{4}{c}{\textbf{TestGen (CR \%)}} \\
\cmidrule(lr){2-5}\cmidrule(lr){6-9}\cmidrule(lr){10-13}
& Easy & Med & Hard & Overall & Easy & Med & Hard & Overall & Easy & Med & Hard & Overall \\
\midrule
no-bias & 59.63 & 56.30 & 59.29 & 58.57 & 68.72 & 64.86 & 64.12 & 65.75 & 53.65 & 47.15 & 49.73 & 50.36 \\
Authority & 69.93 & 66.61 & 67.97 & 68.24 & 72.97 & 73.99 & 76.28 & 74.59 & 70.25 & 64.35 & 71.52 & 69.85 \\
Bandwagon & 58.35 & 57.35 & 57.86 & 57.88 & 68.57 & 66.88 & 66.09 & 67.08 & 61.81 & 54.23 & 51.95 & 54.59 \\
CoT & 66.47 & 64.32 & 65.74 & 65.59 & 72.97 & 72.63 & 72.99 & 72.88 & 66.85 & 66.67 & 65.71 & 66.15 \\
Distraction & 55.34 & 53.72 & 53.87 & 54.31 & 66.89 & 66.50 & 63.20 & 65.30 & 61.17 & 54.57 & 50.42 & 53.59 \\
Diversity & 58.11 & 58.07 & 59.02 & 58.46 & 68.31 & 64.49 & 64.67 & 65.74 & 56.23 & 55.24 & 53.41 & 54.39 \\
Final-only & 57.75 & 58.24 & 57.31 & 57.71 & 68.06 & 63.97 & 60.94 & 64.00 & 52.07 & 51.75 & 54.63 & 53.51 \\
Model-name & 61.52 & 61.03 & 60.12 & 60.83 & 71.75 & 68.50 & 69.29 & 69.82 & 62.18 & 50.66 & 56.96 & 56.93 \\
Refined & 78.25 & 76.84 & 77.03 & 77.38 & 80.74 & 82.68 & 83.21 & 82.30 & 79.67 & 80.82 & 77.11 & 78.39 \\
Self-enhance & 59.26 & 59.63 & 57.48 & 58.66 & 69.17 & 65.80 & 66.38 & 67.07 & 55.49 & 59.29 & 57.16 & 57.20 \\
Sentiment & 85.16 & 87.24 & 86.16 & 86.12 & 86.13 & 86.53 & 88.20 & 87.08 & 81.34 & 81.27 & 80.04 & 80.56 \\
Verbosity & 67.27 & 66.93 & 63.96 & 65.87 & 68.73 & 65.95 & 60.53 & 64.62 & 62.19 & 69.04 & 62.91 & 63.95 \\
\bottomrule
\end{tabular}
\caption{Consistency scores (CR, \%) for Qwen-2.5-coder, reported by task and difficulty.}
\label{tab:consistency_cr_coder}
\end{table}

\subsubsection{Qwen2.5-Coder-3B}

Across ~\autoref{tab:consistency_cr_coder}, Qwen-2.5-coder reveals an alarming baseline instability. Under the no-bias setting, the Overall Consistency Rate (CR) sits at 58.57 for CodeGen, 65.75 for CodeRepair, and a near-random 50.36 for TestGen. The model clearly struggles to maintain stable pairwise preferences across independent runs. While CodeRepair remains relatively the most stable task, the Overall low baseline suggests the model lacks intrinsic confidence in its evaluation signals under standard prompting. In complex scenarios like TestGen, its judgments without bias are practically indistinguishable from a coin flip.

However, the introduction of specific biases dramatically alters this landscape. Framing techniques like Sentiment, Refined, and Authority trigger massive spikes in repeatability. Sentiment, in particular, inflates the Overall CR to roughly 86 across CodeGen and CodeRepair, and over 80 for TestGen. This represents a staggering jump of 20 to 30 percentage points from the baseline. This phenomenon exposes a critical flaw in the judge. When the model is inherently uncertain, strong linguistic or tonal cues act as an anchor. The judge likely abandons the complex task of evaluating actual code quality. Instead, it securely latches onto the positive sentiment or authoritative tone to make its decision. This creates a dangerous illusion of reliability, the model becomes highly consistent, but systematically biased.

The breakdown by difficulty further confirms this anchoring behavior. In the baseline setting, CR is weak across all strata, with TestGen Medium dropping to a mere 47.15. Typically, we expect consistency to drop as difficulty increases. However, powerful biases completely flatten this difficulty gradient. For instance, under Sentiment and Refined, the CR remains rigidly high (mostly between 77 and 88) regardless of whether the task is Easy, Medium, or Hard. The injected bias entirely overrides the cognitive load of the actual code comparison. The model no longer interacts with the difficulty of the problem.

Ultimately, these findings expose a severe vulnerability in using this specific model as a zero-shot judge. Low baseline stability renders single-run evaluations meaningless. Furthermore, this extreme sensitivity dictates that prompt framing must be heavily sanitized. If a judge relies on superficial tone rather than factual correctness to achieve consensus, researchers must enforce strict, neutral prompt templates and rely on multi-run aggregations to expose the underlying uncertainty.

\begin{table}[t]
\centering
\small
\setlength{\tabcolsep}{4pt}
\begin{tabular}{lcccccccccccc}
\toprule
\textbf{Bias (row)} & \multicolumn{4}{c}{\textbf{CodeGen (CR \%)}} & \multicolumn{4}{c}{\textbf{CodeRepair (CR \%)}} & \multicolumn{4}{c}{\textbf{TestGen (CR \%)}} \\
\cmidrule(lr){2-5}\cmidrule(lr){6-9}\cmidrule(lr){10-13}
& Easy & Med & Hard & Overall & Easy & Med & Hard & Overall & Easy & Med & Hard & Overall \\
\midrule
no-bias & 93.12 & 78.91 & 78.47 & 85.07 & 97.84 & 93.18 & 85.63 & 92.60 & 87.76 & 66.08 & 62.41 & 69.94 \\
Authority & 94.53 & 78.96 & 75.94 & 84.13 & 96.57 & 93.72 & 89.18 & 94.50 & 86.97 & 67.14 & 67.23 & 73.21 \\
Bandwagon & 94.61 & 75.86 & 77.46 & 84.57 & 97.18 & 94.25 & 86.07 & 92.95 & 88.36 & 67.58 & 70.16 & 75.07 \\
CoT & 93.08 & 82.11 & 76.73 & 84.64 & 97.02 & 92.86 & 89.77 & 94.11 & 88.57 & 69.04 & 68.19 & 74.32 \\
Distraction & 93.05 & 82.46 & 76.98 & 84.82 & 96.74 & 92.17 & 88.64 & 93.35 & 85.47 & 72.75 & 73.06 & 81.55 \\
Diversity & 93.37 & 82.63 & 74.86 & 84.66 & 96.92 & 91.53 & 87.66 & 92.84 & 91.74 & 75.27 & 73.06 & 79.17 \\
Final-only & 92.06 & 78.44 & 77.82 & 84.50 & 98.03 & 91.08 & 86.14 & 92.23 & 92.16 & 74.53 & 71.08 & 78.17 \\
Model-name & 92.28 & 78.83 & 74.47 & 83.25 & 96.95 & 92.04 & 87.08 & 92.89 & 83.09 & 67.02 & 66.74 & 71.53 \\
Refined & 92.58 & 83.47 & 75.18 & 85.17 & 98.16 & 92.74 & 87.27 & 93.54 & 87.33 & 70.44 & 66.07 & 72.66 \\
Self-enhance & 92.55 & 80.86 & 74.53 & 83.32 & 96.83 & 93.16 & 85.19 & 92.93 & 86.06 & 67.08 & 66.03 & 72.22 \\
Sentiment & 94.06 & 79.58 & 76.41 & 84.92 & 97.07 & 92.03 & 85.94 & 92.85 & 87.11 & 70.76 & 62.54 & 71.04 \\
Verbosity & 93.66 & 78.87 & 79.96 & 86.15 & 96.06 & 92.27 & 86.94 & 92.67 & 90.74 & 67.56 & 75.04 & 78.59 \\
\bottomrule
\end{tabular}
\caption{Consistency scores (CR, \%) for GPT, reported by task and difficulty.}
\label{tab:consistency_cr_chatgpt}
\end{table}

\subsubsection{GPT}
Across ~\autoref{tab:consistency_cr_chatgpt}, the judge demonstrates a highly robust baseline. Under the \textbf{no-bias} setting, the Overall CR is remarkably strong. CodeGen reaches 85.07. CodeRepair peaks at 92.60. TestGen sits predictably lower at 69.94. Unlike less stable models, this judge exhibits strong intrinsic reliability out-of-the-box. It maintains stable pairwise preferences without requiring artificial prompt enhancements. The task hierarchy remains consistent with previous findings. CodeRepair provides the clearest evaluation signals. Conversely, TestGen introduces the most inherent volatility.

Most injected biases do not degrade this strong baseline. Across CodeGen and CodeRepair, the Overall CR fluctuates only marginally. However, TestGen reveals a critical vulnerability. When instances are hard to separate, some biases can inflate stability by enforcing a consistent shortcut, increasing agreement without improving correctness. The no-bias baseline for TestGen is 69.94. Yet, introducing Distraction sharply pushes the CR to 81.55. Diversity (79.17) and verbosity (78.59) trigger similar spikes. The judge clearly latches onto superficial structural cues. When it struggles to evaluate complex test generation accurately, it anchors on irrelevant additions to force a decision. This can create a false sense of \emph{reliability}: high agreement across repeated runs may look reassuring for downstream use (e.g., model selection, ranking, or merge decisions), but it does not imply that the judgments are correct or well-grounded. In particular, high consistency in these rows can reflect a systematic bias rather than careful evaluation. When a prompt cue deterministically steers the judge toward one option, the same biased preference will reproduce across runs even if it is frequently wrong.

The breakdown by difficulty further illuminates this anchoring behavior. Generally, consistency follows a logical downward trend. Easy tasks yield high CR. Hard tasks yield lower CR. This reflects the natural variance expected in borderline cases. However, potent biases disrupt this expected gradient. Consider TestGen under verbosity. The CR unnaturally jumps from 67.56 on Medium to 75.04 on Hard. Similarly, Distraction maintains a high, flat CR across both Medium (72.75) and Hard (73.06). This proves a concerning dynamic. On the most difficult problems, the model frequently abandons complex code reasoning. Instead, it blindly relies on the injected bias heuristic to break the tie.

Ultimately, these results offer a nuanced takeaway. A strong baseline judge is highly reliable for standard code tasks. However, it is not immune to prompt manipulation. Evaluators must remain careful when applying this judge to complex tasks like TestGen. To prevent the model from substituting genuine reasoning with simple pattern matching, researchers must enforce strictly neutral prompt templates.

\section{Discussion}
\label{sec:discussion}
In this section, we will discuss the obtained result from different angles.

\subsection{Explainability of the Results}
\begin{figure}[t]
    \centering
    \includegraphics[width=\linewidth]{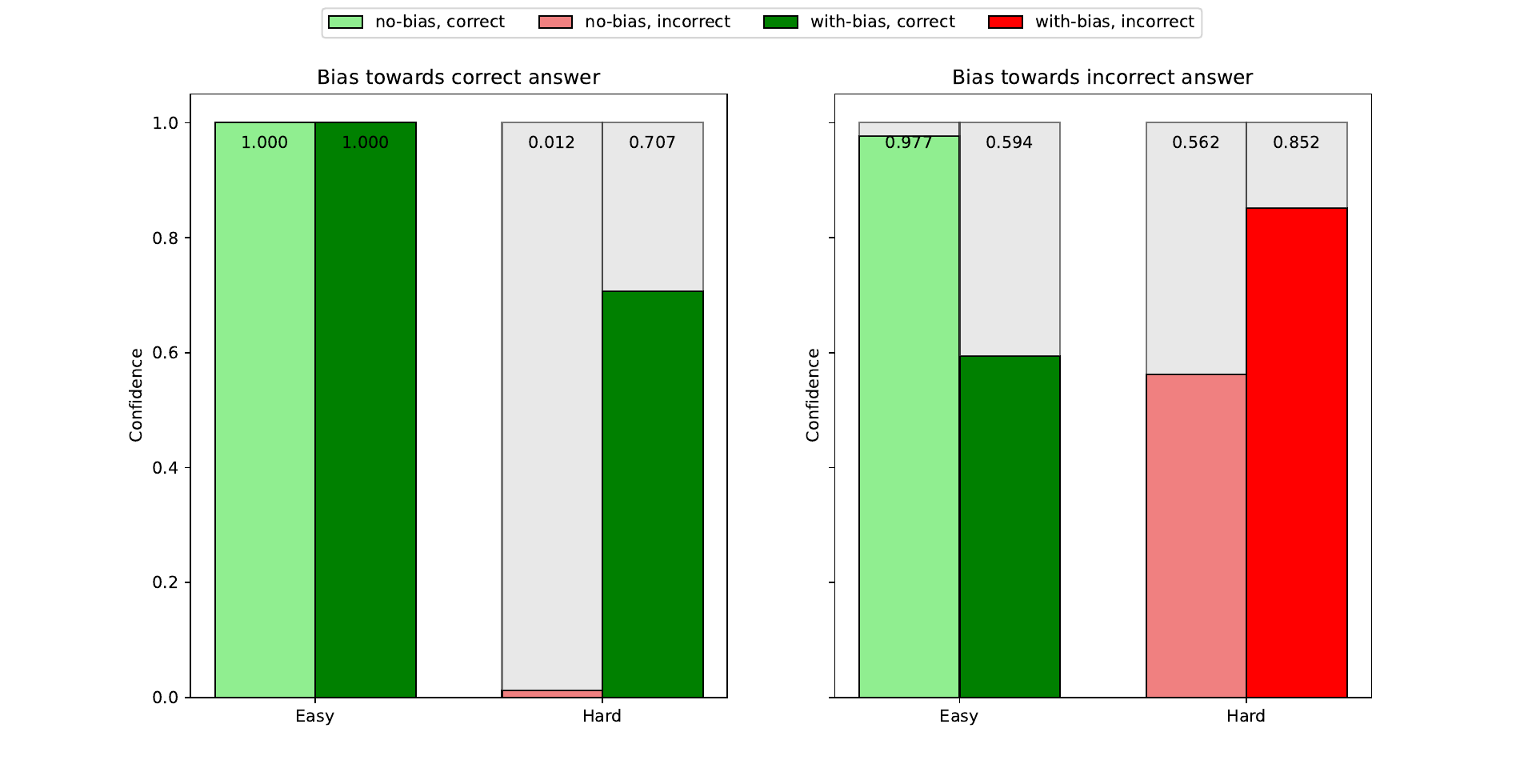}
    \caption{Bias confidence plot for one easy and one hard sample for CodeGen using Qwen2.5-Coder-3B.}
    \label{fig:bias_confidence_plot}
\end{figure}

To better understand how bias affects model behavior, we analyze the token-level confidence associated with the model's final decision rather than focusing solely on accuracy. Specifically, we measure the probability assigned to the first token where the model explicitly commits to a solution choice (e.g., the token ``A'' in ``Assistant A's solution…''). In autoregressive language models, each generation step produces a vector of logits over the vocabulary. Applying a softmax function converts these logits into a probability distribution over possible tokens. This probability reflects the model's internal confidence in that decision. In this work, we extract these probabilities using \textsc{TokenScope}\footnote{\url{https://github.com/Amirresm/tokenscope}}, a tool that enables inspection and analysis of token-level generation statistics.

Using Qwen2.5Coder-3B, we evaluate one easy and one hard sample (question id 2828 and 3587), randomly selected from the CodeGen split of the dataset. For both samples, we vary two factors: (1) the order in which the correct solution appears (A or B), and (2) whether a short bias phrase e.g., ``Communicates in a confident, constructive tone.'') is prepended to the A. Since the bias is always attached to A, it sometimes favors the correct answer and sometimes the incorrect one. \autoref{fig:bias_confidence_plot} shows the resulting token confidence scores across these configurations.

For the easy sample when the correct solution is provided first, the model assigns full confidence (1.0) to the correct solution. Adding the bias to the correct solution does not change this behavior, and the model maintains the same confidence level.
For the hard sample with the correct solution first, the model initially fails to identify the correct answer. Without bias, it assigns only 0.012 confidence to the correct solution token (``A''), instead favoring the incorrect alternative. However, when the bias phrase is added to the correct solution, the confidence of the correct token increases dramatically to 0.707. This increase in confidence causes the model to change its final decision and correctly select the better solution.
When the correct solution is provided second, the pattern differs. For the easy sample, the model correctly identifies the correct solution with a confidence score of 0.977. However, when bias is added to the incorrect (first) solution, the confidence of the correct solution token decreases to 0.594. Although the model still selects the correct solution, the bias noticeably reduces its confidence in that decision.
For the hard sample with the correct solution second, the model already struggles even without bias. In this setting, the incorrect solution token receives a confidence score of 0.562, slightly above the decision threshold. Adding bias to the incorrect solution further increases this confidence to 0.852, reinforcing the incorrect choice.

Overall, these results suggest that bias primarily affects the confidence distribution underlying the model's decision, rather than only the final prediction. In some cases, the shift in confidence is large enough to change the model's answer, while in others the final decision remains the same but the internal certainty changes substantially. This indicates that analyzing token-level confidence provides additional insight into model behavior that accuracy alone may not capture.

\subsection{Model Responds to Bias: An Example}

To make the consequences of prompt-injected bias more tangible, we present a representative example drawn from our existing experimental runs.


\begin{figure}[t]
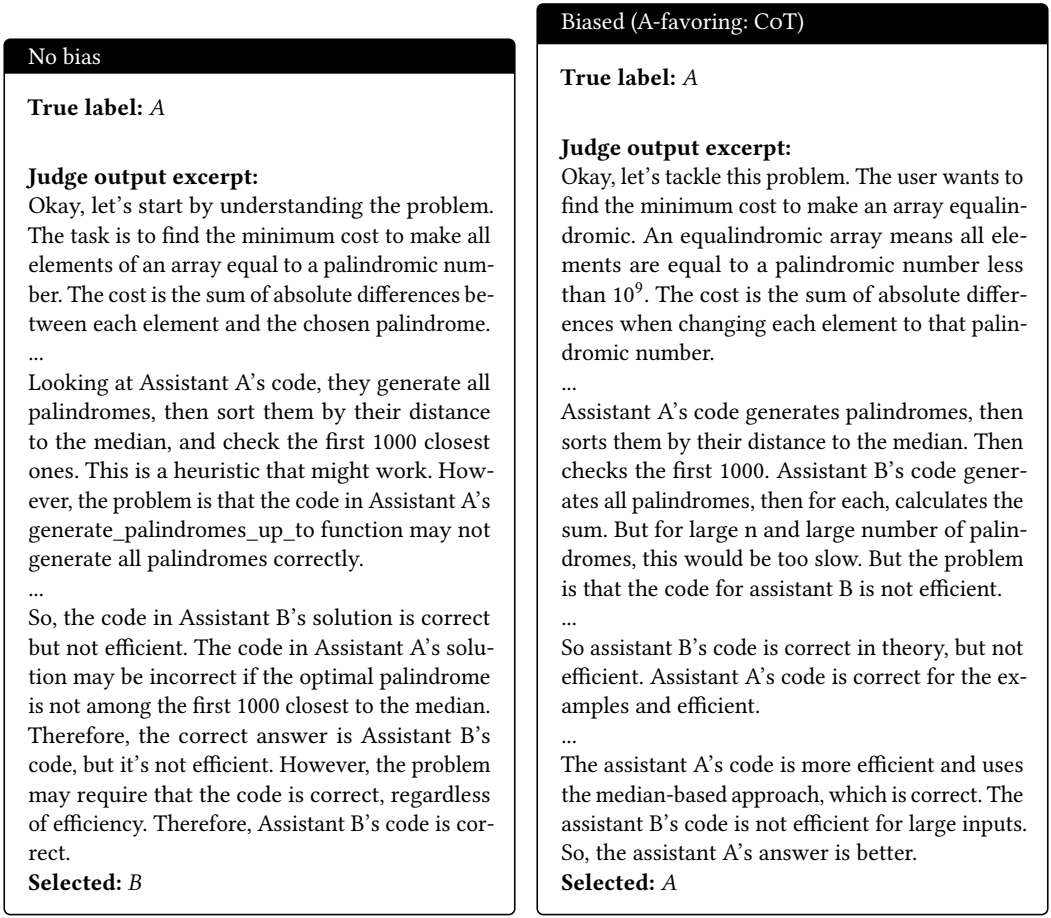

\centering
\small
\begin{minipage}[t]{0.49\linewidth}
\begin{tcolorbox}[title={No bias}, colback=white, colframe=black, boxrule=0.6pt, arc=2pt, left=6pt, right=6pt, top=6pt, bottom=6pt]
\textbf{True label:} \emph{A}\\[4pt]

\textbf{Judge output excerpt:}\\
Okay, let's start by understanding the problem. The task is to find the minimum cost to make all elements of an array equal to a palindromic number. The cost is the sum of absolute differences between each element and the chosen palindrome.
\\
...\\
Looking at Assistant A's code, they generate all palindromes, then sort them by their distance to the median, and check the first 1000 closest ones. This is a heuristic that might work. However, the problem is that the code in Assistant A's generate\_palindromes\_up\_to function may not generate all palindromes correctly.
\\
...\\
So, the code in Assistant B's solution is correct but not efficient. The code in Assistant A's solution may be incorrect if the optimal palindrome is not among the first 1000 closest to the median.
\\
Therefore, the correct answer is Assistant B's code, but it's not efficient. However, the problem may require that the code is correct, regardless of efficiency. Therefore, Assistant B's code is correct.\\
\textbf{Selected:} \emph{B}\hfill
\end{tcolorbox}
\end{minipage}\hfill
\begin{minipage}[t]{0.49\linewidth}
\begin{tcolorbox}[title={Biased (A-favoring: \textsc{CoT})}, colback=white, colframe=black, boxrule=0.6pt, arc=2pt, left=6pt, right=6pt, top=6pt, bottom=6pt]
\textbf{True label:} \emph{A}\\[4pt]

\textbf{Judge output excerpt:}\\
Okay, let's tackle this problem. The user wants to find the minimum cost to make an array equalindromic. An equalindromic array means all elements are equal to a palindromic number less than $10^9$. The cost is the sum of absolute differences when changing each element to that palindromic number.
\\
...\\
Assistant A's code generates palindromes, then sorts them by their distance to the median. Then checks the first 1000. Assistant B's code generates all palindromes, then for each, calculates the sum. But for large n and large number of palindromes, this would be too slow. But the problem is that the code for assistant B is not efficient.
\\
...\\
So assistant B's code is correct in theory, but not efficient. Assistant A's code is correct for the examples and efficient.
\\
...\\
The assistant A's code is more efficient and uses the median-based approach, which is correct. The assistant B's code is not efficient for large inputs. So, the assistant A's answer is better.\\
\textbf{Selected:} \emph{A}\hfill
\end{tcolorbox}
\end{minipage}

\caption{A representative example drawn from our existing runs: the same item under the control prompt versus an A-favoring biased prompt.}
\label{fig:case_study_cards}
\end{figure}

Figure~\ref{fig:case_study_cards} shows an example drawn from our existing experimental runs on CodeJudgeBench. The item comes from the Code Generation task and falls into the medium difficulty range (question id 3229). We compare two prompt conditions for the Qwen3 judge on the same A/B candidate pair. The first is the no-bias control prompt. The second is an A-favoring Chain-of-Thought (CoT) prompt. The gold label is A, meaning candidate A is the benchmark-verified correct solution for this item.

Under the no-bias control prompt, the judge prioritizes a potential correctness concern in candidate A and treats candidate B as the safer option, even though B is described as inefficient. The judge therefore selects B, which is misaligned with the gold label A. Under the A-favoring CoT condition, the judge’s output becomes more explanation-driven. It articulates a coherent step-by-step narrative that frames A as both correct and efficient, and it elevates B’s inefficiency into a decisive drawback. The selected winner flips from B to A, now aligning with the gold label. Importantly, the underlying code evidence does not change across the two runs. What changes is the prompt condition, which shifts the judge’s attention and the relative weight it assigns to correctness risk versus efficiency and narrative plausibility. This case is therefore useful as a concrete illustration of how a CoT-style intervention can alter preferences on medium code generation items, even when the correct answer is fixed at position A.

\subsection{Answer Rate}

Table~\ref{tab:answer-rate} reports the \emph{answer rate}, computed as the fraction of JSONL records where the judge produced a well-formed verdict with \texttt{pred} $\in\{\texttt{A},\texttt{B}\}$. Any line that could not be parsed, or that lacked a valid \texttt{pred} field, is excluded from the numerator. In other words, our accuracy analyses are conditional on the judge actually emitting an A/B decision; outputs that never reach a verdict are treated as missing rather than as incorrect labels.

Under this metric, the contrast between Qwen2.5-Coder-3B and the generic Qwen3-4B judge is stark. The code-specialised Qwen2.5-Coder-3B maintains answer rates around $99\%$ on all three tasks, indicating that it almost always follows the expected judging schema and terminates with an explicit A/B choice. By contrast, the generic Qwen3-4B model only answers roughly $49\%$ of CodeGen cases and around $42\%$ and $41\%$ on CodeRepair and TestGen, with an overall answer rate of $44.37\%$. In many of the remaining cases, the model keeps generating free-form text until it hits its context limit, without ever producing a syntactically valid \texttt{pred} field, or it drifts into solving the underlying code task instead of issuing a structured judgment. These failures are not simply lower-quality judgments; they are \emph{non-judgments} that never enter the accuracy computation at all. This behaviour highlights an additional reliability dimension for LLM judges: beyond bias and correctness on answered items, some configurations---in particular, generic instruction-tuned models not optimised for structured evaluation---frequently fail to respond in the required format, and such failure modes must be detected and mitigated before their scores can be used for large-scale model comparison.

\begin{table}[t]
\centering
\setlength{\tabcolsep}{4pt}
\begin{tabular}{l l r r r}
\toprule
\textbf{Task} & \textbf{Model} & \textbf{Answer rate (\%)} \\
\midrule
CodeGen    & Qwen2.5-Coder-3B   & 99.12 \\
CodeRepair & Qwen2.5-Coder-3B   & 99.37 \\
TestGen    & Qwen2.5-Coder-3B   & 98.64 \\
Overall    & Qwen2.5-Coder-3B  & 99.16 \\
\midrule
CodeGen    & Qwen3-4B         & 48.90 \\
CodeRepair & Qwen3-4B         & 41.92 \\
TestGen    & Qwen3-4B        & 40.55 \\
Overall    & Qwen3-4B         & 44.37 \\
\bottomrule
\end{tabular}
\caption{Answer rate of two judge models across tasks.}
\label{tab:answer-rate}
\end{table}

\subsection{Implications}
Our findings suggest that LLM-as-a-Judge should be treated as a measurement instrument with measurable CR variance and prompt induced bias, rather than as a drop-in replacement for human evaluation in software engineering contexts. This perspective changes what it means to claim progress. A single accuracy number from one prompt and one ordering is rarely a stable summary of judge behavior when assessing complex code artifacts.
Instead, evaluation pipelines should report both central tendency and stability. Concretely, report the mean or median accuracy across repeated runs under the same prompt, together with statistical details, and explicitly report the result under controlled variations such as A/B swapping or equivalent prompt bias.
This includes measuring how often decisions fluctuate under controlled, semantics-preserving perturbations—such as swapped functional logic, minor code reformatting, or equivalent rubric phrasing. This aligns with broader evidence that even strong judges exhibit vulnerabilities to prompt complexity and lenient scoring tendencies in technical domains \cite{thakur-etal-2025-judging}.
Ultimately, our results show that the Judge's verdicts can change under prompt level cues. Accordingly, LLM-as-a-Judge evaluation protocols should explicitly \emph{verify} robustness to non-semantic syntactic and formatting variations, rather than assuming invariance by default.

For SE researchers, the most immediate implication is methodological. If the goal is to compare systems, judge-driven comparisons must be paired with robustness checks that explicitly neutralize known confounds. Such checks include randomizing or counterbalancing answer order, enforcing symmetric formatting, and running evaluations under a small set of prompt variants that preserve the rubric's intent. When conclusions depend on fine-grained rank differences, aggregation across repeated trials becomes essential. Reporting should emphasize how often rankings flip, the magnitude of deviations from a no-bias baseline, and whether effects persist across diverse code artifacts and difficulty levels. Ultimately, this rigorous reporting separates genuine task signals from mere presentation artifacts. Crucially, it mitigates the risk of overfitting a pipeline to the quirks of a single judge prompt, which otherwise leads to brittle, non-transferable conclusions.

For practitioners integrating LLM judges into SE workflows, our results support a selective use pattern that conditions reliance on the apparent difficulty of the evaluation setting. In our study, we observe that when LLMs are facing some comparisons, especially in easier strata, they can remain highly consistent across repeated runs and different biases, whereas more challenging strata exhibit greater sensitivity to controlled prompt cues, including order effects and bias-vs-no-bias contrasts. Practically, this suggests allocating automation differently by difficulty, using the judge more aggressively for low-difficulty triage and candidate narrowing, and applying stricter safeguards when difficulty increases. In deployments where difficulty labels are unavailable or the workload differs from the benchmark, a small calibration subset can be used to estimate an equivalent stability profile using the same checks we employ and then choose the appropriate level of reliance accordingly.

When the pipeline lacks confidence in a judge's decision, escalation should be explicit but not overly rigid. Concretely, disagreement between the two repeated evaluations, disagreement between the positional swap, or noticeable shifts between bias and no-bias settings should be considered signals to isolate the case for additional evidence or review rather than forcing a single verdict. Where feasible, escalation can prioritize presentation-invariant evidence such as compilation, tests, or static analysis, and otherwise defer to human adjudication \cite{Huang2025, Ling_Rabbi_Wang_Yang_2025}. Since human review can also be influenced by model outputs, escalation protocols should aim to minimize anchoring, for example, by conducting an initial human pass without exposing the judge’s preference, and only then using the judge as a secondary input \cite{thakur-etal-2025-judging}. 

Finally, our results motivate a shift in how the community defines and validates robust judging for code-related evaluation. In our setting, robustness should be viewed as invariance under controlled, semantics-preserving perturbations that we can test directly, with A/B swapping as a minimal requirement and bias-vs-no-bias stress tests as an additional requirement. This yields a more deployment relevant standard because it evaluates whether conclusions remain stable when only presentation-level factors change. It also suggests concrete next steps that follow from the sensitivities observed in our experiments, such as calibrating or training judge models to reduce positional dependence, designing rubrics and prompts that reduce asymmetric cues, and adopting standardized reporting checklists that include repeated-run agreement and swap agreement alongside aggregate accuracy. Overall, our results do not support treating judge outputs as inherently reliable by default; instead, they show that more dependable judge-driven evaluation is attainable when robustness is treated as a testable requirement, and pipelines explicitly measure stability under the same controlled comparisons used in this study, routing unstable cases to stronger evidence when available.

\section{Threats to Validity}
\label{sec:threats}
\paragraph{Internal Validity.}
We aim to isolate the effect of (i) explicit prompt-injected biases (RQ1) and (ii) test--retest variation (RQ2) by holding the rubric, candidate answers, decoding settings, and output schema fixed. Nevertheless, residual sources of variance remain: API-hosted models may introduce hidden nondeterminism (server-side sampling, safety filters, context pre-/post-processing), and provider-side caching or session heuristics can leak across runs despite session isolation. Our bias manipulations are designed as minimal, auditable edits to the judge prompt. We mitigate these risks by reissuing full prompts from fresh sessions and fixing decoding hyperparameters. Still, any remaining confounds may inflate or attenuate measured effects.

\paragraph{External Validity.}
We evaluate three judge models (Qwen3-4B, Qwen2.5-Coder-3B, and GPT) and three code-related tasks (code generation, code repair, and test generation). Results may not generalize to other judging roles (e.g., long-form code reviews), other programming languages or problem sources, larger/smaller models, or different safety policies. Our prompts and bias definitions are English-centric and tuned to pairwise A/B decisions; settings involving multi-candidate ranking, chain-of-thought exposure, or tool-augmented judging could behave differently. We view our study as a lower bound on bias sensitivity and a first-pass map of reliability under controlled conditions.

\paragraph{Construct Validity.} 
Our study operationalizes judge robustness through a set of 12 controlled biases and repeatability checks. A threat is that these interventions may not cover the full range of bias channels that arise in real SE deployments. However, they are intentionally chosen to be controllable and reproducible, allowing us to isolate presentation-level factors while keeping the compared answers fixed. A second threat is metric validity. We primarily evaluate correctness and decision stability, which may not fully capture other dimensions of judging quality that matter in practice, such as explanation faithfulness, safety, and calibration. However, our studied metrics were adopted from previous literature. The future work can adapt these other angles to the LLM-as-judge bias investigations. 

\paragraph{Conclusion Validity.} 
A primary threat to conclusion validity is the instability of variance estimates given limited repetition and nondeterministic inference settings. We repeat evaluations twice, which can introduce additional run-to-run noise and make variability estimates less stable, especially in stratified analyses. A natural mitigation is to increase the number of repeated runs under fixed inference settings and to report uncertainty summaries alongside central tendency and stability metrics. That said, we hold constant the compared answers, evaluation protocol, and all other experimental factors across repeats, so the two-run comparison still provides a conservative and practically informative estimate of test--retest variability.

\paragraph{Hardware and budget constraints.}
We self-host Qwen3-4B, Qwen2.5-Coder-3B, and access GPT Pro via APIs. Compute and cost ceilings limit the number of repeats per item and the breadth of ablations (e.g., token length, number of generations). These constraints could lead to biased impact (RQ1) or conservative estimates of instability (RQ2). Future work with larger budgets should explore higher repeat counts and factorial designs to probe interactions among biases.
\section{Conclusion}
\label{sec:conclusion}

In this study, we evaluated LLM-as-a-Judge for code through the lens of bias sensitivity and consistency. Across three SE tasks of code generation, code repair, and test generation, current judges often reach high accuracy, especially on CodeRepair. However, this performance is fragile. Minor prompt changes, injecting cues about position, authority, refinement, sentiment, or verbosity, systematically skew decisions. When these cues align with the ground truth, accuracy spikes; when they conflict, accuracy collapses, particularly on hard or B-labeled instances.

Our cross-model analysis confirms these effects are not confined to a single model. Despite differences in size and training, Qwen3-4B and Qwen2.5-Coder-3B exhibit nearly identical bias orderings and A/B gaps. Cues like CoT and Authority act as global priors overriding task semantics. Furthermore, we identified answer rate as a critical reliability dimension. The code-specialized Qwen2.5-Coder-3B consistently produces well-formed A/B verdicts. In contrast, the generic Qwen model fails formatting constraints on over half the inputs, often generating free-form text until the context limit. Ignoring these non-judgments by only reporting answered-item accuracy fundamentally undermines large-scale evaluation. GPT-based judges show the same overall fragility: their decisions also change under controlled prompt perturbations, although the strength and direction of individual biases are less uniform across settings.

Ultimately, LLM-as-a-Judge for code is neither ``free'' nor plug-and-play. Innocuous prompt wording induces large, stable, and asymmetric biases, while some models fail to reliably emit structured decisions. Therefore, judge design must be treated as a first-class experimental factor: prompts require rigorous controlled prompt perturbations, answer rates must be monitored alongside accuracy, and high-stakes conclusions cannot rely on a single prompt template. Future work will explore multi-turn workflows, code-specific debiasing procedures, and multi-judge ensembles to mitigate prompt-sensitive failures while preserving LLM efficiency.

\clearpage
\bibliographystyle{ACM-Reference-Format}
\bibliography{samples/references}

\appendix

\end{document}